\documentclass[aps,showpacs,prb,amsmath,twocolumn,floatfix]{revtex4}

\usepackage{bm}
\usepackage{epsf}
\usepackage{array}
\usepackage{epsf}
\usepackage{array}

\def\tK{T_{\rm K}} 
\def\ns{n_{\rm s}}   
\def\vF{v_{\rm F}}   
\def\bfp{{\bf p}}    
\def\taue{\tau_{\rm e}}  
\def\vare{\varepsilon}
\def\taus{\tau_{\rm s}}  
\def\Tsg{T_{\rm sg}}  
\def\tTsg{\tilde T_{\rm sg}}  

\usepackage{epsf}
\usepackage{array}
\usepackage{bm}
\begin{document}
\title{Electron transport and energy relaxation in dilute magnetic alloys}
\author{M. G. Vavilov,$^1$ L. I. Glazman,$^1$ and A. I. Larkin$^{1,2}$}
\affiliation{$^1$ Theoretical Physics Institute, University of Minnesota,
Minneapolis, MN 55455\\
$^2$ Landau Institute for Theoretical Physics, Moscow, 117940 Russia}
\date{May 9, 2003}

\begin{abstract}
We consider the effect of the RKKY interaction between magnetic
impurities on the electron relaxation rates in a normal metal. The
interplay between the RKKY interaction and the Kondo effect may result
in a non-monotonic temperature dependence of the electron momentum
relaxation rate, which determines the Drude conductivity.
The electron phase
relaxation rate, which determines the magnitude of the weak
localization correction to the resistivity, is also a non-monotonic
function of temperature. For this function, we find the dependence of
the position of its maximum on the concentration of magnetic
impurities. We also relate the electron energy relaxation rate to the
excitation spectrum of the system of magnetic impurities. The energy
relaxation determines the distribution function for the
out-of-equilibrium electrons. Measurement of the electron distribution
function thus may provide information about the excitations in the
spin glass phase.
\end{abstract}

\maketitle

\section{Introduction}
\label{sec:1}

Electron transport in normal metals is known to be very sensitive to
the presence of magnetic impurities in a metal. Scattering of
conduction electrons off such impurities scrambles the electron spin.
A tiny concentration of magnetic impurities results in an observable
effect -- low-temperature saturation of the phase relaxation
rate~\cite{Pierre}.  Magnetic impurities apparently also facilitate
the energy transfer between electrons~\cite{PierreJLTP,KG}. At higher
concentrations~\cite{Pierre}, a minimum in the temperature dependence
of the resistivity becomes evident, which is a manifestation of the
Kondo effect~\cite{Kondomin}. These three observations fit very well
with a picture of uncorrelated magnetic impurities.

Investigation of the resistivity at even higher concentration of
magnetic impurities, reveals deviations from the picture of
uncorrelated localized magnetic moments. The temperature dependence of
the resistivity, in addition to the aforementioned minimum, develops a
maximum~\cite{Laborde,Ford,mydosh,larsen,larsen1} at lower temperatures.
The electron phase relaxation rate was recently
measured~\cite{MohWebb,Bauerle} on
AuFe alloys with magnetic impurity (Fe) concentration ranging from
$7.1$ to $60$~ppm. At the upper end of this range, correlations
between the localized spins may become important, as evidenced by the
temperature dependence of the resistivity~\cite{Laborde,Bauerle}.

In this paper we investigate the effect of correlation between
spins of magnetic impurities on the electronic transport properties of a
metal.  Specifically, we study the Kondo contribution to the Drude
resistivity, the weak-localization correction to the conductivity, and
the electron energy relaxation rate.

The leading mechanism of correlations is known to be the
Ruderman-Kittel-Kasuya-Yosida (RKKY) interaction between impurities.
The interaction lifts the degeneracy in the excitation spectrum of
impurities. In this respect, the effect of RKKY interaction is
somewhat similar to the effect of an external magnetic field which causes
Zeeman splitting of the spin states. It is known that the magnetic
field reduces the electron relaxation rates at low
energies~\cite{KG,BFK,VG,Goppert}, and also suppresses the Kondo
effect~\cite{KondovsMagField}.  Unlike the uniform Zeeman splitting,
however, interaction between the spins results in a broad spectrum of
energies of collective spin states. Therefore, the quantitative
manifestations of the RKKY interaction in the electron transport are
different from that of the Zeeman energy.

The difficulty of the low-temperature electron transport problem is
associated with the complexity of the spin glass state.  In this
paper, we study in detail the transport at relatively high
temperatures (or electron energy transfers, in the case of energy
relaxation).  We perform analytical calculation using the method of
the virial expansion\cite{LMK,LKh} in the RKKY interaction between
magnetic impurities. The method is based on the following concept.
Since the RKKY interaction decreases fast as a function of the
distance between impurities, the impurities have to be close to each
other for the interaction between them to compete with thermal
smearing and to affect transport properties of conduction electrons.  We
perform the virial expansion to the second order in the density of
magnetic impurities, which corresponds to accounting for the
interaction within impurity pairs.

Electron scattering off magnetic impurities contributes to the
temperature dependence of the resistivity: due to the Kondo effect,
the resistivity increases as the temperature is lowered. Similar to
the Zeeman splitting, the RKKY interaction between magnetic impurities
may stop the development of the Kondo effect. The interplay between
the RKKY interaction and the Kondo effect leads to a maximum in the
temperature dependence of the resistivity.\cite{mydosh,larsen,larsen1}
If the
characteristic temperature of the spin glass formation  exceeds
significantly the
Kondo temperature (high concentration of magnetic impurities $\ns$),
then the resistivity has a maximum at a temperature
in the region of applicability of the virial expansion.  On the
contrary, at small $\ns$ the maximum of resistivity occurs at zero
temperature.

Magnetic impurities also affect the magneto-resistance in weak
magnetic fields. This magneto-resistance is due to the weak
localization (WL) effect.\cite{HLN,WL} Being an interference
phenomenon, the WL is limited by the electron phase relaxation.
We calculate the phase relaxation rate taking into account the RKKY
interaction between magnetic impurities. At high impurity concentration, the
phase relaxation rate has a maximum, similar to the maximum in the
resistivity. At low concentration, unlike the Drude resistivity,
the phase relaxation retains a maximum at a finite temperature
(of the order of the Kondo temperature).

The RKKY interaction lifts the degeneracy of the magnetic impurity
states.  Therefore an out-of-equilibrium electron may loose its energy
by exciting the impurity spin degrees of freedom. The corresponding
relaxation rate is a function of the transferred energy. This rate
provides information about the spin excitations spectrum. We
derive the corresponding kinetic equation for the electron
distribution function. We also make specific predictions for the
relaxation rate at sufficiently large energy transfers, which may be
accounted for by the virial expansion.

Before we proceed, we emphasize that in this paper we consider the effect
of the interaction between Kondo impurities only on
kinetic properties (conductance, energy relaxation)
of electrons in a metal.
The interaction between magnetic impurities may also affect
thermodynamic properties (such as heat capacity, susceptibility,
superconducting transition temperature)
of conduction electrons.\cite{LKh,LMK,GL,BA}
An analysis of the thermodynamic
properties of electrons in dilute Kondo alloys was performed
earlier, see e.g. ref.~\onlinecite{BA}.

The paper is organized as follows. In the next section we introduce the model
and discuss the effect of interaction between  magnetic impurities
on a spin correlation function.
In Sec. III we perform calculations of the resistivity correction due to the
electron scattering off magnetic impurities. Section IV contains analysis of
the WL correction to the conductivity. In Sec. V we derive the kinetic
equation for the system of magnetic impurities and conduction electrons
and discuss the energy exchange rate between these two subsystems.
Section VI contains discussions and conclusions.

\section{Model}
\label{sec:2}

The scattering of conduction electrons off a magnetic impurity is described by
the following Hamiltonian
\begin{equation}
\hat H_{\rm e}={\cal J}\hat{\bm{S}}\hat{\bm{\sigma}},
\label{eq:He}
\end{equation}
where $\hat{\bm{S}}$ is the spin operator of a magnetic impurity and $\hat{\bm \sigma}$
is the spin operator of a conduction electron represented in terms of the Pauli
matrices $\{\hat \sigma_x,\hat \sigma_y,\hat \sigma_z\}$.
The exchange constant ${\cal J}$ is renormalized due to the Kondo effect and
varies as a logarithmic function of energy $\vare$ of conduction
electrons. At temperature $T$ higher than the Kondo temperature $\tK$,
the exchange constant for thermal electrons, $\vare\sim T$, is
given by
\begin{equation}
{\cal J}=\frac{2}{\nu}\ln^{-1}\frac{T}{\tK}, \ \ \  T\gg\tK,
\label{eq:JKondo}
\end{equation}
where $\nu$ is the Fermi density of states per spin degree of freedom.

In order to evaluate the effects of electron scattering off magnetic
impurities on electron transport, we introduce the statistical averages
of impurity spin components, $\langle \hat S_\alpha \hat S_\beta\rangle$, with
respect to thermodynamic states of the magnetic impurity system. We show that
the electron transport properties in a
metal with strong spin-orbit coupling are determined by the following spin
correlator
\begin{equation}
K(t)=\langle \hat{\bm{S}}(0)\hat{\bm{S}}(t)\rangle.
\label{eq:Kdef}
\end{equation}
In general, the correlation function $K(t)$ can be rewritten
in terms of exact quantum
states $|\xi\rangle$ of the system of magnetic impurities:
\begin{equation}
K(\omega) = 2\pi \sum_{\xi\xi'}\rho_\xi
\left|
\langle \xi | {\bm S} | \xi'\rangle
\right|^2
\delta(E_\xi-E_{\xi'}-\omega).
\label{eq:Kgeneral}
\end{equation}
Here $E_\xi$ is the energy of state $|\xi\rangle$, and $\rho_\xi$ is
the density matrix $\rho_\xi\propto \exp (-E_\xi/T)$.

For a free magnetic impurity with spin $S$ (all spin states are degenerate)
the spin correlation function $K(t)$ does not depend on time and its Fourier transform
has the form:
\begin{equation}
K_{1}(\omega)=2\pi S(S+1)\delta(\omega).
\label{eq:K0}
\end{equation}
Using the Fermi golden rule and Eq.~(\ref{eq:K0}), we obtain the
following expression for the electron scattering rate off magnetic impurities
\begin{equation}
\frac{1}{\taus}=2\pi \nu n_{\rm s}{\cal J}^2S(S+1).
\label{eq:1taus}
\end{equation}
Here $n_{\rm s}$ is the magnetic impurity concentration per
volume. The quantity $\taus$ is the mean free time of scattering off magnetic
impurities.

In metals, the leading interaction between magnetic moments of impurities
is described by the Ruderman-Kittel-Kasuya-Yosida (RKKY) mechanism. The
corresponding Hamiltonian has the form:
\begin{equation}
\hat H_{\rm RKKY}=\sum\limits_{ij}V(r_{ij})\hat{\bm{S}}_i\hat{\bm{S}}_j.
\label{eq:HRKKY}
\end{equation}
The magnitude of the RKKY interaction is given by the following expression
\begin{equation}
V(r)=\frac{V_0(r)}{r^3}\cos\varphi,
\label{eq:VRKKY}
\end{equation}
where $\varphi$ changes fast on the length scale of the Fermi wavelength
$\lambda_{\rm F}$.
The interaction constant $V_0$ may be represented in terms of
the exchange constant and electron density of states $\nu$:
\begin{equation}
V_0(r)=\frac{\nu {\cal J}_0^2(r)}{2\pi}, \ \ \
{\cal J}_0(r)=\frac{2}{\nu}\frac{1}{\ln [v_{\rm F}/(r\tK)]}.
\label{eq:V0}
\end{equation}
The typical value of the RKKY interaction between two impurities
separated by distance $1/\ns^{1/3}$ (average distance between
impurities) is
\begin{equation}
\Tsg= \ns V_0(r=\ns^{-1/3})\simeq\frac{2\ns}{\pi\nu\ln^2(v_{\rm F}\ns^{1/3}/\tK)}.
\label{eq:Tsg}
\end{equation}

Due to the randomness of the RKKY Hamiltonian $\hat H_{\rm RKKY}$,
finding of states $|\xi\rangle$ and corresponding energies
$E_\xi$ is a hardly possible task.  At sufficiently low temperature
a phase transition into a spin glass state may occur with extremely
complicated structure of the wave functions $|\xi\rangle $
and energy spectrum.\cite{FischerHertz}
A transition temperature is comparable with
the typical energy of the interaction between magnetic impurities $\Tsg$.

We focus our attention on the high temperature limit $T\gg\Tsg$
and calculate the spin-spin correlation function $K(t)$,
Eq.~(\ref{eq:Kdef}), using the virial expansion method.
In this method the interaction between magnetic
impurities is taken into account only if the
splitting of the spin states due to the interaction exceeds the system
temperature;
otherwise the interaction does not significantly
change the spin correlation function.
Therefore, the interaction is important only for magnetic impurities
in clusters of the size $\sim \sqrt[3]{V_0/T}$.
For a uniform distribution of magnetic impurities
in the metal, the probability for a formation of such a cluster
of $k$ impurities scales as
$(\Tsg/T)^{k-1}$.  We consider only clusters containing
two ($k=2$) magnetic impurities.

The energy states $|\xi\rangle$ of two interacting spins
are classified by the total spin $J$ ($J=0,1,\dots,2S$) and its projection
$M$ on a fixed direction:
$|\xi \rangle = |J,M \rangle$.
The energy spectrum is given by
\begin{equation}
E_J=V(r)\frac{J(J+1)-2S(S+1)}{2}\equiv V(r) \epsilon_J
\label{eq:EJ}
\end{equation}
and is degenerate with respect to the projection $M$.
The spacing between levels with different $J$
is proportional to the magnitude of the RKKY interaction,
Eq.~\ref{eq:VRKKY}.

According to Eq.~(\ref{eq:Kgeneral}), the corresponding spin correlation
function of a magnetic impurity within distance $r$ from another magnetic
impurity may be represented in the form
\begin{eqnarray}
\!\!\!\!  K_2(t,V) =
\!\!
\sum\limits_{J,J'=0}^{2S}\!\!
A_{JJ'}\frac{ 2J+1 }{Z(V)}
e^{V\left\{i(\epsilon_J-\epsilon_{J'})t-\epsilon_J/T\right\}}
,
\label{eq:deltaK}
\end{eqnarray}
where the statistical sum $Z(V)$ is
\begin{subequations}
\label{eqs:ZA}
\begin{eqnarray}
Z(V) & = & \sum\limits_{J=0}^{2S} (2J+1)e^{-V\epsilon_J/T},
\label{eq:Zvirial}
\end{eqnarray}
and the matrix elements $A_{JJ'}$ are
\begin{equation}
A_{JJ'}=\sum_{M'}\langle JM |\hat{\bm{S}} |J'M'\rangle
\langle J'M' |\hat{\bm{S}} |JM\rangle .
\label{eq:ALL-def}
\end{equation}
\end{subequations}
The analytical form of the matrix elements is
presented in Appendix A. We emphasize that $A_{JJ'}$
are independent of the pair spin projection $M$.

As we will see in the following sections, the electron transport
properties are determined by the spin correlation function
$\overline{K(\omega)}$, averaged over configurations of magnetic
impurities. We calculate $\overline{K(\omega)}$ within the virial
approximation:
\begin{equation}
\overline{K(\omega)}= K_1(\omega)+\int p(r) \left[
K_2(\omega,V(r))-K_1(\omega)
\right]
d^3{\bm r},
\label{eq:barKformal}
\end{equation}
where $p(r)$ is the probability density for two magnetic
impurities to be at distance $r$; for uniform impurity
distribution $p(r)=\ns$. Averaging over the relative
position of two magnetic impurities in Eq.~(\ref{eq:barKformal})
can be performed in two steps. First, we make the substitution
$r^3=V_0\cos\varphi/(Ty)$, and then we perform integration over the
fast varying phase $\varphi$. As the result,
the spin correlator has the form\cite{LMK}
\begin{eqnarray}
\overline{ K(\omega)} & = &
K_1(\omega)+\delta\overline{K_2(\omega)},
\label{eq:barKvirial}
\end{eqnarray}
where
\begin{subequations}
\label{eq:K2}
\begin{eqnarray}
\!\!\!\!\!\!\!\!\! \delta\overline{K_2(\omega)}\!\!\!
&=&\!\!\!
\frac{8\pi}{3}\frac{\Tsg S(S+1)}{T}\!\!
\int\limits_{-\infty}^{+\infty}
\!\!
\frac{dy}{y^2}\left[
 P(\omega,y)-\delta(\omega)
\right],
\label{eq:deltaK2}
\\
\!\!\!\!\!\!\!\!\!P(\omega,y) & =&
\!\! \sum\limits_{JJ'=0}^{2S}\!\!A_{JJ'}\frac{2J+1}{S(S+1)}
\frac{e^{-y\epsilon_J}}{Z(T y)}
\nonumber
\\
&&\times
\delta(\omega-T y[\epsilon_{J}-\epsilon_{J'}]).
\label{eq:P}
\end{eqnarray}
\end{subequations}

We notice that the spin correlation function
Eq.~(\ref{eq:barKvirial}) increases as the frequency $\omega$
decreases:
\begin{eqnarray}
\overline{ K(\omega)} & = &
\frac{8\pi}{3}\frac{\Tsg}{T}
\sum\limits_{J\neq J'}^{2S}(2J+1)A_{JJ'}
\frac{|\epsilon_J-\epsilon_{J'}|}{\omega^2}
\nonumber
\\
&& \times
\frac{\displaystyle \exp\left(-\frac{\omega\epsilon_J}
{(\epsilon_J-\epsilon_{J'})T}\right)}
{Z(\omega/(\epsilon_J-\epsilon_{J'}))}
,\ \ \omega\neq 0.
\label{eq:barKnonzero}
\end{eqnarray}
One may expect that due to the $1/\omega^2$ behavior of the spin
correlation function, Eq.~(\ref{eq:barKnonzero}), the virial
approximation breaks down even at  $T\gtrsim \Tsg$. Nevertheless,
due to the property $P(\omega,0)=\delta(\omega)$, see
Eq.~(\ref{eq:sumrule}),
the integrand in Eq.~(\ref{eq:deltaK2}) has no singularity at
$y=0$, and the slow modes ($\omega\lesssim T$) of the spin system
do not affect electron transport.

Equation (\ref{eq:barKvirial}) supplemented  with
Eqs.~(\ref{eqs:ZA}) and (\ref{eq:K2})
determines the spin correlation function at
high temperature $T\gg \Tsg$.
Below we use these equations to describe the effect of interaction
between magnetic impurities on electron transport in metals.

\section{Resistivity}
\label{sec:3}

The conductivity of a metal with isotropic
impurities may be
calculated according to the standard rules of the diagrammatic
technique. Disregarding the interference corrections we have
\begin{equation}
\sigma=\frac{e^2\vF^2}{6}\int
\frac{\overline{G^{R}}(\vare,\bfp)\overline{G^{A}}(\vare,\bfp)}
{T\cosh^2\vare/2T}
\frac{d\vare d\bfp}{(2\pi)^4}.
\end{equation}
Here
\begin{equation}
\overline{G^{R,A}}(\vare,\bfp)=\frac{1}{\vare-\xi(\bfp)-\Sigma^{R,A}(\vare)}
\label{eq:Gra}
\end{equation}
is the retarded or advanced Green function averaged over disorder,
$\xi({\bfp})= v_{\rm F}(|\bfp|-p_{\rm F})$
is the electron energy, counted from the Fermi energy,
$\Sigma^{R,A}(\vare)$ is the electron retarded or advanced
self energy, and $\vF$ is the Fermi velocity.
Performing the integration over momentum $\bfp$, we obtain
\begin{equation}
\sigma=e^2\nu\frac{\vF^2}{3}\int
\frac{1}{{\rm Im}\Sigma^{A}(\vare)}
\frac{d\vare}{4T\cosh^2\vare/2T}
\label{eq:sigma}
\end{equation}
with $\nu$ being the density of states of conduction electrons
per one spin orientation.

We assume that the electron self energy part contains two
components:
\begin{equation}
{\rm Im} \Sigma^{A}(\vare)= \frac{1}{2\taue} +\ns {\rm Im}
\overline{T(\vare)}.
\label{eq:ImSigma}
\end{equation}
The first term in Eq.~(\ref{eq:ImSigma}), $1/2\taue$,
represents the effect of elastic scattering off
non-magnetic impurities with $\taue$ being the mean free
elastic time. The second term,
$\ns{\rm Im}\overline{T(\varepsilon)}$, represents the effect of scattering of
electrons with energy $\varepsilon$ off magnetic impurities.
Here the scattering off a particular magnetic impurity is characterized
by the $T-$matrix $T(\vare)$; the self energy $\Sigma^{R}(\vare)$ contains
$\overline{T(\vare)}$, averaged over various impurities.

Using the simple relation $\rho=1/\sigma$ between the conductivity $\sigma$
and the resistivity $\rho$ and Eqs.~(\ref{eq:sigma}) and
(\ref{eq:ImSigma}),
we represent the resistivity as a sum of two terms
\begin{equation}
\rho=\rho_{\rm e} +\Delta\rho_{\rm K}.
\label{rho}
\end{equation}
The first term is the resistivity of a metal without magnetic impurities
($\ns=0$), which is produced by elastic scattering
off non-magnetic impurities:
\begin{equation}
\rho_{\rm e}=\frac{1}{\sigma_{\rm e}},
\quad
\sigma_{\rm e}=2e^2\nu D,
\end{equation}
where $D=v_{\rm F}^2\taue/3$ is the diffusion coefficient.
The second term is the contribution
to the resistivity due to the scattering off magnetic impurities:
\begin{equation}
\Delta\rho_{\rm K}=
\frac{3\ns}{e^2\nu\vF^2}
\int  {\rm Im}\overline{T(\vare)}
\frac{d\vare}{4T\cosh^2\vare/2T}
\label{eq:deltasigmam}
\end{equation}
The scattering $T-$matrix in Eq.~(\ref{eq:deltasigmam}) has
different structure in the limits of high ($T\gg\tK$)
and low ($T\ll\tK$) temperatures. We study these two limits below.

\subsection{High concentration of magnetic impurities, $\Tsg\gg\tK$}
\label{sec:3A}

At high temperature $T\gg\tK$ the scattering of electrons off magnetic
impurities is described by the Born approximation with the
exchange constant renormalized according to Eq.~(\ref{eq:JKondo}).
In this case the $T-$matrix is (see Appendix \ref{app:B}):
\begin{equation}
{\rm Im}T(\vare)= \pi\nu {\cal J}^2\int K(\omega)
\frac{1+e^{\vare/T}}{1+e^{(\vare-\omega)/T}}
\frac{d\omega}{2\pi},
\label{eq:ImT}
\end{equation}
where $K(\omega)$ is the Fourier transform of the spin correlation
function, defined by Eq.~(\ref{eq:Kdef}).
Substituting ${\rm Im} T(\vare)$ into Eq.~(\ref{eq:deltasigmam}),
we obtain the Kondo contribution to the Drude conductivity:
\begin{equation}
\Delta \rho_{\rm K}=
\frac{3\ns {\cal J}^2}{2 e^2\vF^2}
\int
\overline{K(\omega)}
\frac{\omega}{T}\frac{d\omega}{1-e^{-\omega/T}} .
\label{eq:rhoKgeneral}
\end{equation}
We emphasize that Eq.~(\ref{eq:rhoKgeneral}) is
valid if the distance between impurities is much larger than
the Fermi wavelength. At the same time, Eq.~(\ref{eq:rhoKgeneral})
describes the resistivity in metals with arbitrary
structure and strength of interaction between
magnetic impurities.\cite{FischerHertz} We perform further calculations using
the spin correlation function $\overline{K(\omega)}$ given
by Eq.~(\ref{eq:barKvirial}), which was derived within the virial
expansion.

Using ${\overline K(\omega)}$,
calculated within the virial expansion
Eq.~(\ref{eq:barKvirial}), and the Kondo renormalized exchange constant
${\cal J}$, see Eq.~(\ref{eq:JKondo}), we obtain
\begin{equation}
\Delta \rho_{\rm K} =\frac{12\pi\ns}{e^2v_{\rm F}^2\nu^2}
\frac{S(S+1)}{\ln^2(T/\tK)}
\left(
1-\alpha_S \frac{\Tsg}{T}\right).
\label{eq:rhoKvirial}
\end{equation}
Here numerical coefficient $\alpha_S$ is given by the following integral:
\begin{equation}
\alpha_S=\int\limits_{-\infty}^{+\infty}\!\!\!
\left(
1-\sum\limits_{JJ'}^{2S}\frac{y(\epsilon_J-\epsilon_{J'})e^{-y\epsilon_J}}
{1-e^{-y(\epsilon_J-\epsilon_{J'})}}
\frac{(2J+1)A_{JJ'}}{S(S+1)Z(y)}
\right)
\frac{dy}{y^2},
\label{eq:alphaS}
\end{equation}
where $\epsilon_J$, $Z(y)$ and $A_{JJ'}$ are defined by Eqs.~(\ref{eq:EJ}),
(\ref{eq:Zvirial}) and (\ref{eq:ALL-def}) respectively.
We emphasize that the integral in Eq.~(\ref{eq:alphaS}) converges near $y=0$.
The values of $\alpha_S$ are presented in Table~\ref{tab:1}.

Equation~(\ref{eq:rhoKvirial}) is similar to the results of
Refs.~\onlinecite{LMK,larsen}. Unlike Ref.~\onlinecite{LMK},
Eq.~(\ref{eq:rhoKvirial}) takes into account the Kondo
renormalization of the exchange constant, Eq.~(\ref{eq:JKondo}),
and of the RKKY interaction, Eq.~(\ref{eq:Tsg}).
The advantage of Eq.~(\ref{eq:rhoKvirial}) in comparison
with Ref.~\onlinecite{larsen} is in a consistent definition of
the energy scale $\Tsg$ and in an accurate
procedure of the averaging over states of magnetic impurities, which
allowed us to calculate the numerical factor $\alpha_S$.

\begin{table}
\caption{Values of the numerical coefficient $\alpha_S$,
Eq.~(\ref{eq:alphaS}), for several values of $S$.}
\begin{ruledtabular}
\begin{tabular}{|c|c|c|c|c|c|c|c|}
  $\displaystyle S$ & 1/2  & 1 & 3/2 & 2 & 5/2 & 3 & 7/2 \\
  \hline
  $\alpha_S$ & 1.10 & 1.52 & 1.85 & 2.11 & 2.33 & 2.52 & 2.69 \\
\end{tabular}
\end{ruledtabular}
\label{tab:1}
\end{table}

In metals with low Kondo temperature, $\tK\lesssim\Tsg$,
the competition between the Kondo effect and the effect of
RKKY interaction results in a maximum of the
resistivity as a function of temperature. If $\tK\ll \Tsg$, then
the maximum occurs at
\begin{equation}
T^*\simeq \frac{\alpha_S}{2}\Tsg\ln\frac{\Tsg}{\tK},
\label{eq:Tstarv}
\end{equation}
see also Ref.~\onlinecite{mydosh,larsen,larsen1}.
We notice that temperature $T^*$ is
within the region of applicability of the virial expansion, used in
the derivation of Eq.~(\ref{eq:rhoKvirial}).

As temperature $T$ approaches and crosses $\Tsg$, intrinsic random
magnetic field develops, and the renormalization of the exchange
constant is stopped at ${\cal J}\simeq 2/(\nu\ln
(\Tsg/\tK))$. Simultaneously, the virial expansion breaks down, and
the collective modes of spin system have to be considered in the
derivation of the temperature dependence of the resistivity.

At present, there is no theory of metallic spin glasses, which
would provide us with $\overline{K(\omega)}$ in a broad range of
$\omega$. Thus, the explicit form of $\Delta\rho_{\rm K}(T)$, see
Eq.~(\ref{eq:rhoKgeneral}), is not known. We expect that
$\Delta\rho_{\rm K}$ continues to decrease with temperature $T$ decreasing,
as the dynamics of the local magnetic moments gets progressively
suppressed at lower temperatures. In the mean field picture, each of
the spins at $T=0$ is subject to a finite field, so that such dynamics is
fully suppressed. In this case, the limiting value of
$\Delta\rho_{\rm K}$ at $T=0$ can be estimated as
\begin{equation}
\Delta \rho_{\rm K}(T=0) =\frac{12\pi\ns}{e^2v_{\rm F}^2\nu^2}
\frac{S^2}{\ln^2(\Tsg/\tK)}
\label{eq:rhoKSG}
\end{equation}
(The quenching of the spin flips leads to the replacement of $S(S+1)$
factor, see Eq.~(\ref{eq:rhoKvirial}), by the factor $S^2$ here.) The
picture leading to Eq.~(\ref{eq:rhoKSG}) can not be valid for the
``tightest'' pairs of magnetic moments with the characteristic
interaction energy significantly exceeding $\Tsg$. It is not clear to
us at the moment even in which direction the estimate
(\ref{eq:rhoKvirial}) changes due to the deviations from the mean
field description.

\subsection{Low concentration of magnetic impurities, $\Tsg\ll\tK$}
\label{sec:3B}

At high temperature ($T\gg\tK$), the resistivity is
described by Eq.~(\ref{eq:rhoKvirial}) even in the limit
$\tK\gg\Tsg$. Nevertheless,  the effect of interaction between
magnetic impurities is small due to the factor $\Tsg/T$ in
Eq.~(\ref{eq:rhoKvirial}), and the maximum of the resistivity
does not occur at  $T>\tK$.
We show that at temperatures $T\lesssim\tK$ the resistivity
also monotonically increases as temperature decreases.

For this purpose we use the following form of the imaginary
part of the scattering $T-$matrix:\cite{AL}
\begin{equation}
{\rm Im}T(\vare) = \frac{1}{\pi\nu}-{\rm Im}\tilde T(\vare).
\label{eq:TMatsubU}
\end{equation}
Here the first term is the contribution to the $T-$matrix in the
unitary limit and
$\tilde T(\vare)$ is the $T-$matrix, written for the residual
interaction between conduction electrons and magnetic impurities.
The form of $\tilde T(\vare)$ differs for spin impurity
$S=1/2$ (exactly screened magnetic impurities) and
for $S>1/2$ (underscreened magnetic impurities).\cite{AL}
We consider the two cases in more details below.

\subsubsection{Spin $S=1/2$ impurities}
\label{sec:3B1}

In this case the majority of
magnetic impurities are completely screened and
the RKKY type interaction between them is absent
(only a small part $\propto \Tsg/\tK$ of magnetic impurities
form coupled states with binding energy exceeding $\tK$). Therefore
the contribution to the resistivity is determined
by the $T-$matrix of a single magnetic impurity,
which is well studied at $T\ll\tK$.\cite{NB,AL}
We present the results for
convenience. The scattering matrix for the residual interaction
between conduction electrons and magnetic impurities has the form:
\begin{equation}
{\rm Im}\tilde T(\vare)=\frac{1}{\nu}
\frac{9\pi}{8\tK^2}\left(3\vare^2+\pi^2T^2
\right).
\label{eq:Texactlyscreened}
\end{equation}
Substituting Eqs.~(\ref{eq:TMatsubU}) and (\ref{eq:Texactlyscreened})
into Eq.~(\ref{eq:deltasigmam}) and performing integration over energy
$\vare$, we obtain:\cite{AL}
\begin{equation}
\Delta\rho_{\rm K}=\Delta\rho_{\rm U}\left[1-
\frac{9\pi^4}{4}\frac{T^2}{\tK^2}
\right].
\label{eq:rhoKSGU}
\end{equation}
Here the factor
\begin{equation}
\Delta\rho_{\rm U}=\frac{3}{\pi}\frac{\ns}{e^2\nu^2v_{\rm F}^2},
\label{eq:rhoU}
\end{equation}
corresponds to the unitary
contribution to the resistivity at $T=0$.
According to Eq.~(\ref{eq:rhoKSGU})
at finite temperature the resistivity contains corrections to
Eq.~(\ref{eq:rhoU}), which are proportional to $T^2/\tK^2$.

We also notice that the coefficients in Eq.~(\ref{eq:rhoKSGU}) contain
small corrections $\Tsg/\tK$ due to magnetic
impurities, which form coupled states with binding energies exceeding
the Kondo temperature.
Two possibilities exist: i) if the coupled state is a singlet,
these impurities do not affect electron transport; ii) if the coupled state
is a triplet, it becomes screened and again leads to a $T^2$
dependence of the resistivity on temperature.\cite{Varma} We emphasize
that the ratio of the number of such impurities to the total number of
magnetic impurities is small as $\Tsg/\tK$.

As the impurity concentration $\ns$ increases, and consequently $\Tsg$
increases, the system of magnetic impurities with spin $S=1/2$
undergoes a quantum phase transition~\cite{Hertz} to a spin glass
state.

\subsubsection{Impurities with $S>1/2$}
\label{sec:3B2}

As was shown in ref.~[\onlinecite{NB}], at $T\ll\tK$ the residual
coupling of conduction electrons with magnetic impurities
is described by the exchange Hamiltonian
Eq.~(\ref{eq:He})
with effective impurity spin, $\tilde S= S-1/2$, and the renormalized
exchange constant
\begin{equation}
\tilde {\cal J}=\frac{2}{\nu\ln\tK/T}.
\label{eq:JKondoU}
\end{equation}
This coupling results in the RKKY-like interaction
between magnetic impurities, which may be written in the form
of the Hamiltonian given by Eq.~(\ref{eq:HRKKY}) with
the effective spin $\tilde S$ and the strength $V(R)$ of
the RKKY interaction defined as a solution of the following equation:
\begin{equation}
V(R)=\frac{2}{\pi\nu R^3}\frac{1}{\ln^2[\tK/V(R)]}.
\label{eq:Tsgprimeeq}
\end{equation}
Using Eq.~(\ref{eq:Tsgprimeeq}) we estimate the typical value of the
interaction between magnetic impurities.  Within logarithmic accuracy
we have
\begin{equation}
\tTsg \simeq \frac{2\ns}{\pi\nu\ln^2(\tK\nu/\ns)},
\label{eq:Tsgprime}
\end{equation}
cf. Eq.~(\ref{eq:Tsg}).

The imaginary part of the $\tilde T(\vare)-$matrix is given by
Eq.~(\ref{eq:ImT}) with the impurity spin $\tilde S=S-1/2$,
exchange constant ${\cal J}$ given by Eq.~(\ref{eq:JKondoU}),
and the typical value of the RKKY interaction between
impurities $\tTsg$ given by Eq.~(\ref{eq:Tsgprime}). Substituting
${\rm Im}\tilde T(\vare)$ from Eq.~(\ref{eq:ImT}) with the
modified parameters into Eq.~(\ref{eq:TMatsubU}) and using
Eqs.~(\ref{eq:rhoKgeneral}) and (\ref{eq:rhoU}), we obtain
\begin{equation}
\Delta\rho_{\rm K}=\Delta\rho_{\rm U}\left[1-
4\pi^2\frac{S^2-1/4}{\ln^2\tK/T}
\left(
1-\alpha_{S-1/2}\frac{\tTsg}{T}
\right)
\right].
\label{eq:rhoKvirialU}
\end{equation}

Unlike the previously analyzed case of high impurity concentration,
$ T_{\rm sg}\gg \tK$, here the temperature dependence
of the resistivity remains monotonic. The leading contribution to
the (negative) derivative $d\rho/dT$ in the temperature interval
$\tilde T^*\lesssim T\lesssim \tK$ comes from the Kondo
renormalization of the exchange constant; here the characteristic
temperature $\tilde T^*$ is
\begin{equation}
\tilde T^*\simeq \frac{\alpha_{S-1}}{2}\tTsg \ln \frac{\tK}{\tTsg}.
\end{equation}
Below $T^*$, the derivative $d\rho/dT$ is determined by the interaction
between magnetic impurities.

As temperature approaches $\tTsg$, the virial expansion ceases to be
valid, and the system may attain a spin glass state. Unlike the case
of higher impurity concentration ($\Tsg\gg \tK$), here we expect the
dependence $\rho (T)$ to level off at $T\lesssim T_{\rm sg}$. This
difference stems from the behavior of the scattering matrix, see
Eq.~(\ref{eq:TMatsubU}), in the vicinity of the unitary limit. If each
of the local moments is quenched individually, then the saturation would
occur at
\begin{equation}
\Delta\rho_{\rm K}=\Delta\rho_{\rm U}\left[1-
4\pi^2\frac{(S-1/2)^2}{\ln^2\tK/\Tsg}\right].
\label{eq:rhoKSGU11}
\end{equation}
The collective modes existing in the spin glass state would
result, however, in deviations from Eq.~(\ref{eq:rhoKSGU11}).

To summarize Section~\ref{sec:3}, the Kondo contribution to the Drude resistivity
is a non-monotonic function of temperature only in
the case of relatively high concentration of the magnetic impurities,
{\it i.e.}, at $\Tsg\gtrsim\tK$. In the opposite case, this
contribution increases monotonically with the decrease of temperature,
at any value of the spin $S$ of magnetic impurities.

\section{Phase Relaxation Rate}
\label{sec:4}

Weak magnetic fields suppress the interference
contribution to the conductivity, the difference
\begin{equation}
\Delta\sigma_{\rm wl} =
\sigma(B=0)-\sigma(B\gg B_{\rm o})
\label{eq:deltasigmawldef}
\end{equation}
in the
conductivity at zero magnetic field $B=0$ and at sufficiently strong
magnetic field $B\gg B_{\rm o}$ is called the weak localization
correction to the conductivity. The characteristic value $B_{\rm o}$
of magnetic field, which suppresses the weak localization is
\begin{equation}
B_{\rm o}=\frac{\hbar c}{e}\sqrt{\frac{1}{D\taus A}},
\end{equation}
where $A$ is the cross-section area of the wire. Throughout this
Section, we assume that the scattering off magnetic impurities
dominates over all other mechanisms of the electron phase
relaxation.

We calculate $\Delta\sigma_{\rm wl}$
of a metal with magnetic impurities and strong
spin orbit coupling. In this case only the singlet component
${\cal C}_s$ of the Cooperon remains finite,\cite{HLN}
all other Cooperon components are suppressed
by the spin-orbit interaction. The weak (anti)localization
correction to the
conductivity is given by:\cite{AAK,AAG}
\begin{equation}
\Delta\sigma_{\rm wl}=\frac{e^2D}{\hbar}
\int\frac{d\vare d\vare_1d\vare_2}{(2\pi)^3}\frac{d^d{\bm q}}{(2\pi)^d}
\frac{
{\cal C}^{+}_{s}\left(
\begin{matrix}
\vare,\vare_1 \cr
\vare_2,\vare
\end{matrix}
; {\bm q}\right)
}
{2T\cosh^2\vare/2T}.
\label{eq:wlgeneral}
\end{equation}

Neglecting the RKKY interaction between magnetic impurities, we
have the following expression for the Cooperon:
\begin{equation}
{\cal C}^{\pm}_{s,0}\left(
\begin{matrix}
\vare_1,\vare_2 \cr
\vare'_1,\vare_2'
\end{matrix}
; {\bm q}\right)
=
\frac{4\pi^2\delta(\vare_1-\vare_2)\delta(\vare'_1-\vare'_2)}
{D{\bm q}^2\mp i(\vare_1-\vare'_1)+2/\taus},
\label{eq:Coop0}
\end{equation}
where $1/\taus$ is the
electron scattering rate off magnetic impurities and
is defined by Eq.~(\ref{eq:1taus}).\cite{HLN}
In calculating the WL correction, we assume that the rate
$1/\taus$ is higher than the Korringa relaxation rate of the
magnetic impurities, see Ref.~\onlinecite{VG} for further
discussion. Performing integration
over the momentum ${\bm q}$ in a one dimensional case, $d=1$ in
Eq.~(\ref{eq:wlgeneral}),
which adequately describes wires of the cross-sectional area
$A\lesssim D\taus$, we obtain
\begin{equation}
\Delta\sigma^{(0)}_{\rm wl}=\frac{e^2}{2\pi\hbar}
\sqrt{\frac{D\taus}{2}}.
\label{eq:wl1d0}
\end{equation}

We study the effect of interaction between spins of magnetic
impurities on the weak localization correction to the
conductivity. As we have already mentioned, the interaction
between magnetic impurities lifts the degeneracy of impurity spin
states, and therefore it is reminiscent to the Zeeman effect of
external magnetic field. The Zeeman splitting of spin states of
magnetic impurities affects the WL correction to the
conductivity,\cite{BFK,VG} if the splitting is larger than either
temperature $T$ or phase relaxation rate $1/\taus$.  Similarly, the RKKY
interaction between two impurities starts to affect the WL correction
if the RKKY interaction strength exceeds $T$ or $1/\taus$.
We calculate the WL correction at temperatures $T$ higher than $\Tsg$
(or $\tTsg$), so that only a small number of magnetic impurity pairs
satisfy this condition, and therefore the virial expansion is applicable.

We notice that because of the RKKY interaction between impurity
spins, the scattering processes may change electron energy and,
particularly, may switch the position of the Cooperon poles in
energy plane with respect to the real axis. These processes result
in mixing of the Cooperon components ${\cal C}^+$ and ${\cal C}^-$
which have different analyticity, see e.g. Eq.~(\ref{eq:Coop0}).
The full equation for the Cooperon is
\begin{eqnarray}
\hat{\cal C}_s\left(\begin{matrix}
\vare_1,\vare_2 \cr
\vare'_1,\vare_2'
\end{matrix};{\bm q}\right)  =
\hat{\cal C}_{s,0}\left(\begin{matrix}
\vare_1,\vare_2 \cr
\vare'_1,\vare_2'
\end{matrix}{\bm q}\right)
+\int\frac{d\vare_3d\vare_3'd\vare_4d\vare_4'}{(2\pi)^4}&&
\nonumber
\\
\ \ \  \times \hat{\cal C}_{s,0}\left(\begin{matrix}
\vare_1,\vare_3 \cr
\vare'_1,\vare_3'
\end{matrix},{\bm q}\right)
\hat \Sigma\left(\begin{matrix}
\vare_3,\vare_4 \cr
\vare'_3,\vare_4'
\end{matrix}\right) \hat{\cal C}_s\left(\begin{matrix}
\vare_4,\vare_2 \cr
\vare'_4,\vare_2'
\end{matrix},{\bm q}\right).&&
\label{eq:CoopEqn}
\end{eqnarray}
The diagonal elements of the $2\times 2$ matrix $\hat{\cal C}_{s}$
are ${\cal C}^\pm_s$.
The matrix
\begin{equation}
\hat{\cal C}_{s,0}(\cdot,{\bm q}) =
\left(%
\begin{matrix}
 \hat{\cal C}^+_{s,0}(\cdot,{\bm q}) & 0 \\
 0 & \hat{\cal C}^-_{s,0}(\cdot,{\bm q}) \\
\end{matrix}%
\right)
\end{equation}
is the Cooperon to the zeroth order in the RKKY interaction,
see Eq.~(\ref{eq:Coop0}), while the self energy
\begin{equation}
\hat\Sigma_s(\cdot) =
\left(%
\begin{matrix}
 \Sigma_s^{\rm rr}(\cdot) & \Sigma_s^{\rm ra}(\cdot) \\
\Sigma_s^{\rm ar}(\cdot) & \Sigma_s^{\rm aa}(\cdot) \\
\end{matrix}%
\right)
\end{equation}
contains the higher-order RKKY contributions.

To evaluate the first term of the virial expansion, we may account for
$\hat \Sigma_s$ by the first-order iteration of the  solution of
Eq.~(\ref{eq:CoopEqn}). The self energy $\hat \Sigma$ must be
calculated up to the first order in the RKKY interaction.
In fact, it is sufficient
to evaluate the upper diagonal element $\Sigma_s^{\rm rr}(\cdot)$ of the
matrix $\hat \Sigma_s(\cdot)$
and write the Cooperon as:
\begin{eqnarray}
\!\!\!\!\!\!\!\!
{\cal C}^+_s\left(
\begin{matrix}
\vare,\vare_1 \cr
\vare_2,\vare
\end{matrix}
; {\bm q}\right)
\!\! &=&\!\!
\frac{1}{D{\bm q}^2+2/\taus}
+
\frac{1}{D{\bm q}^2+2/\taus-i(\vare-\vare_2)}
\nonumber
\\
&\times &\!\!
\Sigma_s^{\rm rr}\left(
\begin{matrix}
\vare,\vare_1 \cr
\vare_2,\vare
\end{matrix}\right)
\frac{1}{D{\bm q}^2+2/\taus-i(\vare_1-\vare)},
\label{eq:CoopVir}
\end{eqnarray}
with (see Appendix \ref{app:C})
\begin{equation}
\begin{split}
\Sigma&_s^{\rm rr} \left(
\begin{matrix}
\vare,&\vare_1 \\
\vare_2,&\vare
\end{matrix}
\right) = - 4\pi^2 \nu \ns {\cal J}^2\int d\omega
\ \frac{(e^{\vare/T}+1)\delta \overline{K_2(\omega)}}
{e^{(\vare-\omega)/T}+1}
\\
&\times  \left\{
\delta(\vare_1+\omega-\vare)\delta(\vare-\vare_2-\omega) +
 \delta(\vare_1-\vare)\delta(\vare-\vare_2) \right\}.
\label{eq:SEvirial}
\end{split}
\end{equation}
Then, the weak localization correction to the conductivity is
obtained by substituting this expression for
${\cal C}^+_s\left(\cdot; {\bm q}\right)$ into Eq.~(\ref{eq:wlgeneral})
and performing integration over momentum ${\bm q}$ and energies.
As we discussed in the previous Section, details
of the structure of the spin
correlation function $\delta\overline{K_2(\omega)}$ in
Eq.~(\ref{eq:SEvirial}) depend on the relation between temperature
$T$, Kondo temperature $\tK$ and the typical energy of interaction
between impurities $\Tsg$. Some of these limits are discussed below.

\subsection{High concentration of magnetic impurities, $\Tsg\gg\tK$}
\label{sec:4A}

At temperatures $T\gg\tK$ the scattering of electrons off
magnetic impurities is described by the Born approximation with
the renormalized exchange constant, Eq.~(\ref{eq:JKondo}).
We substitute $\delta\overline{K_2(\omega)}$
(the first order term in the RKKY
interaction) from Eq.~(\ref{eq:K2}) into Eq.~(\ref{eq:SEvirial}),
and using Eqs.~(\ref{eq:wlgeneral}) and (\ref{eq:CoopVir}), we
obtain the weak localization correction to the conductivity
[see Appendix \ref{app:D} for more details].
We distinguish three temperature domains for the WL correction to the
conductivity $\Delta\sigma_{\rm wl}$.

In the highest of the three domains, $T\gg 2/\taus$, the weak
localization correction has the form
\begin{equation}
\Delta\sigma_{\rm wl}=
\frac{e^2}{2\pi\hbar}\sqrt{\frac{D\taus}{2}}\left(
1+\frac{\pi(4S+1)(4S+3)}{120(2S+1)}\Tsg\taus \right),
\label{eq:sigmaWLht}
\end{equation}
with the second term in the parentheses coming from the RKKY
interaction. The use of Eq.~(\ref{eq:Tsg}) for $\Tsg$ and of the estimate
for the Kondo-renormalized electron spin relaxation rate,
\begin{equation}
\frac{1}{\taus}=\frac{8\pi\ns}{\nu}
\frac{S(S+1)}{\ln^2T/\tK},
\label{eq:1tausKondo}
\end{equation}
allows us to estimate $\Tsg\taus$ as
\begin{equation}
\Tsg\taus=\frac{1}{4\pi^2S(S+1)}
\frac{\ln^2 (T/\tK)}{\ln^2 (v_{\rm F}\ns^{1/3}/\tK)}.
\label{eq:Tsgtaus}
\end{equation}
We see now that the correction due to the RKKY interaction
only weakly depends on temperature, and is numerically small.

It is curious to notice that the second term in
Eq.~(\ref{eq:sigmaWLht}) is  almost
independent of $\ns$.
This term takes into account the fact that the contribution to the phase
relaxation rate is suppressed, if a scattering process results
in energy exchange larger than $1/\taus$.
The reduction of the phase relaxation rate leads to
the enhancement of the weak localization correction to the
conductivity, as shown in Eq.~(\ref{eq:sigmaWLht}).
The number of impurities with
the splitting of energy states larger than $1/\taus$
constitute only $\Tsg\taus$ part of the
total number of magnetic impurities.     We emphasize that
the accidental numerical smallness of the RKKY-induced correction
$\propto \Tsg\taus$ justifies the use of the conventional theory~\cite{HLN}
of the weak localization in the presence
of magnetic impurities in the considered temperature domain.

In the second temperature domain $\Tsg\lesssim T\lesssim
1/\taus$, the WL correction equals
\begin{eqnarray}
\Delta\sigma_{\rm wl} & = &
\frac{e^2}{2\pi\hbar}\sqrt{\frac{D\taus}{2}}
\left(1+\frac{\alpha_S}{2}\frac{\Tsg}{T} \right)
\nonumber
\\
&\propto &
\ln\frac{T}{\tK}
\left(1+\frac{\alpha_S}{2}\frac{\Tsg}{T} \right).
\label{eq:sigmaWLlt}
\end{eqnarray}
Here the numerical factor $\alpha_S$ is defined in
Eq.~(\ref{eq:alphaS}), and $\Tsg$ is given by Eqs.~(\ref{eq:Tsg}).
The result shown in Eq.~(\ref{eq:sigmaWLlt}) has a similar structure
to the expression for the resistivity correction
Eq.~(\ref{eq:rhoKvirial}). The dependence $\Delta\sigma_{\rm wl}$
vs. $T$ has a minimum at temperature $T^*$ defined in
Eq.~(\ref{eq:Tstarv}). This minimum results from the competition
between two opposite trends: with the reduction of temperature,
$\taus$ gets shorter, see Eq.~(\ref{eq:1tausKondo}), while the
stronger-bound impurity pairs stop affecting $\Delta\sigma_{\rm wl}$.
Note that due to the relation between $\Tsg$ and $1/\taus$, see
Eq.~(\ref{eq:Tsgtaus}), the second temperature domain is rather wide.

The third temperature domain corresponds to the spin glass state of
the magnetic impurities. With temperature decreasing to $\Tsg$, the
virial correction becomes large and Eq.~(\ref{eq:sigmaWLlt}) is no
longer applicable.  At such temperature a spin glass transition is
expected.  Below the transition, $\Delta\sigma_{\rm wl}$ is still
determined by the spin correlation function,
Eq.~(\ref{eq:Kdef}). Similar to the discussion of the resistivity in
Section~\ref{sec:3}, we expect a monotonic increase and saturation of
the WL correction. The limiting value of $\Delta\sigma_{\rm wl}$ at
$T=0$ was estimated in Ref.~\onlinecite{Bergmann}, where quenching of the
dynamics of {\it each} of the local moments was assumed. Deviations
from such a simple picture of the spin glass state would result in a
different value of $\Delta\sigma_{\rm wl}(T=0)$.

\subsection{Low concentration of magnetic impurities, $\Tsg\ll\tK$}
\label{sec:4B}

At $T\gg\tK$, the weak localization correction to the conductivity is
still given by Eq.~(\ref{eq:sigmaWLlt}). However, now the effect of
the RKKY interaction on $\Delta\sigma_{\rm wl}$ is small, and the WL
correction to the conductivity decreases monotonically with the
decrease of temperature.

At temperature $T\ll \tK$ the scattering off a single impurity
approaches the unitary limit. The potential scattering characterizing
the unitary limit, does not destroy phase coherence and thus does
not affect the WL correction. Therefore, at $T\ll\tK$ only
small deviations from the unitary limit determine $\Delta\sigma_{\rm wl}$.
In this section we show that $\Delta\sigma_{\rm wl}$ increases
monotonically as temperature decreases in the domain $T\ll\tK$;
the details of the temperature dependence are different for
$S=1/2$ and $S>1/2$.

Comparing the behavior of WL correction in the domains of low and
high temperatures, we conclude that the correction must have a
minimum at $T\sim \tK$, assuming that the scattering off magnetic
impurities dominates the electron phase relaxation.

\subsubsection{Spin $S=1/2$}
\label{sec:4B1}

At $T=0$, the spins of magnetic impurities are completely screened and
do not contribute to the phase relaxation of the conduction
electrons. At finite but small temperatures, $T\ll\tK$, the residual
local electron-electron interaction facilitated by local moments
leads to the electron relaxation which affects the Cooperon pole:
\begin{equation}
{\cal C}^+_{s,1/2}
\left(
\begin{matrix}
\vare,\vare_1 \cr
\vare_2,\vare
\end{matrix}
; {\bm q}\right)
=\frac{4\pi^2\delta(\vare-\vare_1)\delta(\vare-\vare_2)}
{D{\bm q}^2-i(\varepsilon_1-\varepsilon)+\tilde\Gamma(\vare)}.
\label{eq:Coop1/2}
\end{equation}
Here the relaxation rate
\begin{equation}
\tilde\Gamma(\vare)=\frac{9\pi}{8}\frac{\ns}{\nu}
\frac{3\vare^2+\pi^2T^2}{\tK^2}.
\label{eq:Gamma12}
\end{equation}
Because the interaction responsible for the relaxation is local, the
typical energy transferred in a scattering event is
$\Delta\varepsilon\sim T$, and therefore
$\tilde\Gamma(\vare)/\Delta\varepsilon\ll T\Tsg/\tK^2\ll 1$.
Under these conditions, the Cooperon relaxation rate
$\tilde\Gamma(\vare)$
is just twice the one-electron relaxation rate.

Substituting ${\cal C}^+(\cdot,q)$ from Eq.~(\ref{eq:Coop1/2})
into Eq.~(\ref{eq:wlgeneral}), we obtain
\begin{equation}
\Delta\sigma_{\rm wl}=\frac{e^2}{2\pi\hbar}
\int
\sqrt{
\frac{D}{\tilde\Gamma(\vare)}}\frac{d\vare}{4T\cosh^2\vare/2T}.
\label{eq:wl12}
\end{equation}
According to Eq.~(\ref{eq:wl12}), in the absence of other phase
relaxation mechanisms, the weak localization correction would
vary as $1/T$ at $T\lesssim \tK$:
\begin{equation}
\Delta\sigma_{\rm wl}\approx 0.022
\frac{e^2}{\hbar}\sqrt{\frac{\nu D}{\ns}}\frac{\tK}{T}.
\label{eq:tauphi12}
\end{equation}

We notice, that similar to Eq.~(\ref{eq:rhoKSGU}), there are small
corrections of the order of $\Tsg/\tK$ to the numerical coefficient in
Eq.~(\ref{eq:tauphi12}). The corrections originate from the rare
configurations of ``tight'' pairs of magnetic impurities, which form
singlet or triplet states with binding energy exceeding
the Kondo temperature.

\subsubsection{Spin $S>1/2$}
\label{sec:4B2}

The residual coupling
between conduction electrons and magnetic impurities with $S>1/2$
is still described by the exchange Hamiltonian Eq.~(\ref{eq:He})
with the reduced spin operator $\tilde S=S-1/2$ and the renormalized
exchange constant $\tilde {\cal J}$, see Eq.~(\ref{eq:JKondoU}).
In this case the electron scattering rate is
\begin{eqnarray}
\frac{1}{\tilde\tau_{\rm s}}=
\frac{8\pi\ns}{\nu}\frac{S^2-1/4}{\ln^2\tK/T}.
\label{eq:1tausU}
\end{eqnarray}
The coupling $\tilde {\cal J}$ also results in the RKKY interaction
between the partially screened local moments, which is represented by
the Hamiltonian Eq.~(\ref{eq:HRKKY}) with $\hat{\bf  S}$ replaced by
$\hat{\tilde{\bf  S}}$.  The strength of the RKKY interaction is determined by
the self-consistent equation~(\ref{eq:Tsgprimeeq}).

To calculate the weak localization correction to the conductivity, we
use Eqs.~(\ref{eq:wlgeneral}) and (\ref{eq:CoopVir}) with the self
energy $\Sigma_{\rm rr}$ in the form of Eq.~(\ref{eq:SEvirial}).  The spin
correlation function $\delta \overline{K_2(\omega)}$ in
Eq.~(\ref{eq:SEvirial}) describes correlations of $\tilde S$ spins
with the appropriately replaced exchange constant
Eq.~(\ref{eq:JKondoU}) and the typical value $\tTsg$ of the
RKKY potential, Eq.~(\ref{eq:Tsgprime}).

Similar to Section~\ref{sec:3A}, we can define three domains for the
temperature dependence of the weak localization correction. At high
temperature $T\gtrsim 2/\tilde\tau_{\rm s}$, we obtain
\begin{equation}
\Delta\sigma_{\rm wl} =
\frac{e^2}{2\pi\hbar} \sqrt{\frac{D\tilde\tau_{\rm s}}{2}}
\left[
1+\frac{\pi(16S^2-1)}{240S}\tTsg\tilde\tau_{\rm s}
\right].
\label{eq:sigmaWLUht}
\end{equation}
Equation (\ref{eq:sigmaWLUht}) is a counterpart of
Eq.~(\ref{eq:sigmaWLht}).

At lower temperatures, $\tTsg\lesssim T\lesssim 1/\tilde\tau_{\rm
s}$, we obtain [compare to Eq.~(\ref{eq:sigmaWLlt})]:
\begin{equation}
\Delta\sigma_{\rm wl} =
\frac{e^2}{2\pi\hbar} \sqrt{\frac{D\tilde\tau_{\rm s}}{2}}
\left[
1+\frac{\alpha_{S-1/2}}{2}\frac{\tTsg}{T}
\right].
\label{eq:sigmaWLUlt}
\end{equation}
We notice that at $T\ll \tK$ as temperature decreases the weak
localization correction to the conductivity increases.
We conclude that both the Kondo effect and the effect
of interaction between magnetic impurities reduce the phase relaxation
rate as temperature decreases. At $\tilde T^*\lesssim T\lesssim
\tK$ the temperature dependence of the weak localization correction to
the conductivity is mainly determined by the Kondo effect, and at
lower temperatures, $T\lesssim \tilde T^*$ it is determined by the
interaction between magnetic impurities.  At temperature below $\tTsg$
a spin glass state may appear. Similar to the behavior of
$\Delta\rho_{\rm K}(T)$, the weak localization correction increases
monotonically with the decrease of temperature, and should saturate at
$T\to 0$.

To summarize Section~\ref{sec:4}, the weak localization correction to
the conductivity is a non-monotonic function of temperature.
At relatively high concentration of the magnetic impurities,
{\it i.e.} at $\Tsg\gtrsim\tK$, the positions of minimum in
$\Delta\sigma_{\rm wl}$ and maximum in $\Delta \rho_{\rm K}$
roughly coincide, see Eqs.~(\ref{eq:rhoKvirial}) and
(\ref{eq:sigmaWLlt}). In the opposite case, the minimum in
$\Delta\sigma_{\rm wl}$ occurs at $T\simeq \tK$.

\section{Energy relaxation rate}
\label{sec:5}

Free magnetic impurities are an intermediary for
electron-electron scattering with small energy transfer.\cite{KG}
We show that the RKKY interaction between magnetic impurities
leads to the electron energy relaxation as a result of a single
electron scattering off a magnetic impurity. Indeed, if the impurity
interacts with one or more of its neighbors, a scattering process
is accompanied by the energy exchange between conduction electrons and
magnetic impurities.

In this Section we apply the virial expansion method to derive kinetic
equations for the non-equilibrium distribution function of electrons
in a dilute magnetic alloy.  The virial expansion is justified for
processes with large compared to $\Tsg$ energy transfer from an
electron to the system of localized moments; here $\Tsg$ is the
typical energy of interaction between magnetic impurities, see
Eq.~(\ref{eq:Tsg}).  We assume that the spin-orbit interaction is
strong, in which case the electron distribution is independent of spin
orientation, $f_{\uparrow}(t, {\bm r},\varepsilon_k)=
f_{\downarrow}(t,{\bm r},\varepsilon_k)\equiv f(t, {\bm r},\varepsilon_k)$.

First, we consider electron scattering by magnetic impurities
belonging to a small-size pair. If the electron distribution function
does not significantly vary on the length scale of the order of the pair size,
then the corresponding scattering rate can be expressed in terms of
the electron distribution function $f(t, {\bm r},\varepsilon_k)$ at
the position $\bf r$ of the pair,
\begin{eqnarray}
\Upsilon^{JJ'}_{kk'}(t,{\bm r}) & = &
2\pi {\cal J}^2(2J+1)A_{JJ'}\delta(\vare_k-\vare_{k'}+E_J-E_{J'})
\nonumber
\\
&\times &
P_J(t,{\bm r}, V) f(t,{\bm r},\vare_k)(1-f(t,{\bm r},\vare_{k'})).
\label{eq:UpsilonSR}
\end{eqnarray}
Here $P_J(t,{\bm r}, V)$ is the distribution function for two magnetic
impurities over quantum states characterized by the total spin $J$ of
the pair, $E_J=V\epsilon_J$, see Eq.~(\ref{eq:EJ}), and $A_{JJ'}$ is
defined by Eq.~(\ref{eq:ALL-def}).

Having the rate $\Upsilon^{JJ'}_{kk'}({\bm r})$, we can write the
kinetic equations for the distribution function of the pairs $P_J(V)$.
Performing summation over all initial ($k,\alpha$) and final
($k',\alpha'$) states of a scattered electron as well as over the
final states of the pair, we obtain the following equation:
\begin{equation}
\frac{dP_{J}(t,V)}{dt}=-\frac{4\nu^2}{2J+1}
\sum\limits_{J'}\int d\vare_k d \vare_{k'}
\left(\Upsilon_{kk'}^{JJ'}-\Upsilon_{k'k}^{J'J}
\right)
\label{eq:Prates}
\end{equation}
(we omit the position $\bm r$ of the pair in the argument of $P$). The
normalization condition for $P_J(t,V)$ has the form
\begin{equation}
\sum_{J=0}^{2S}(2J+1)P_J (t,V) =1.
\label{eq:norm}
\end{equation}
In the stationary state, the distribution $P_J(t,V)\equiv P_J(V)$
satisfies the equations
\begin{equation}
P_{J+1}(V)=P_J(V)\frac{\int f(\vare)(1-f(\vare+V\epsilon_{J,J+1}))d\vare}
{\int f(\vare+V\epsilon_{J,J+1})(1-f(\vare)) d\vare},
\label{eq:Pchain}
\end{equation}
where we use the shorthand notation
$\epsilon_{J,J'}=\epsilon_J-\epsilon_{J'}$, and $\epsilon_J$ is
defined by Eq.~(\ref{eq:EJ}). If the system of magnetic impurities
and electrons are at equilibrium with temperature $T$, the solution of
Eq.~(\ref{eq:Pchain}) is the Gibbs distribution:
\begin{equation}
P_J(V)=\frac{\exp(-V\epsilon_J /T)}{\sum_J(2J+1)\exp(-V\epsilon_J /T)}.
\label{eq:Pequil}
\end{equation}

Next, we write the kinetic equation for the electron distribution
function
\begin{equation}
\left[\frac{\partial }{\partial t} -D\frac{\partial ^2 }{\partial^2{\bm r}}
\right] f(t,{\bm r},\vare)=-
{\cal I}(t,\vare_k,V),
\label{eq:kineticeq}
\end{equation}
where the electron collision integral has the form:
\begin{eqnarray}
\!\! {\cal I}(t,\vare_k,V)\!\! &\!\! = & \nu \sum\limits_{JJ'}\int
\left(\Upsilon_{kk'}^{JJ'}-\Upsilon_{k'k}^{J'J}
\right) d\vare_{k'}
\label{eq:Igeneral}
\end{eqnarray}
with $\Upsilon_{kk'}^{JJ'}$ given by Eq.~(\ref{eq:UpsilonSR}).

We assume that the electron distribution function changes slowly with the
coordinate ${\bm r}$, so that the collision integral ${\cal I}$
may be averaged over a small volume of the metal, where
$f(t,{\bm r},\vare_k)$ does not change much, but which
contains many magnetic impurities. In this case we can perform averaging
of the collision integral over the RKKY potential according to:
\begin{equation}
\langle {\cal I}(t,\vare) \rangle=
\frac{4\Tsg}{3}\int\frac{dV}{V^2} {\cal I}(t,\vare,V).
\label{Iavrule}
\end{equation}
Substituting Eq.~(\ref{eq:UpsilonSR}) into
Eq.~(\ref{eq:Igeneral}) and
performing averaging according to Eq.~(\ref{Iavrule}),
we obtain
\begin{widetext}
\begin{eqnarray}
\langle {\cal I}(t,\vare) \rangle  =   \frac{4\Tsg}{3\taus}
\sum\limits_{J\neq J'}
\frac{2J+1}{S(S+1)}A_{JJ'}|\epsilon_{JJ'}|
\int\frac{dE}{E^2}
\left\{
f_{\vare}(1-f_{\vare-E})
P_J\left(t,\frac{-E}{\epsilon_{JJ'}}\right)
-(1-f_{\vare})f_{\vare-E}
P_{J'}\left(t,\frac{-E}{\epsilon_{JJ'}}\right)
\right\},
\label{eq:Iaveraged}
\end{eqnarray}
\end{widetext}
where we use notations $\epsilon_{JJ'}=\epsilon_J-\epsilon_{J'}$ and
$f_\vare=f(t,{\bm r},\vare)$. In the derivation of the collision
integral, we tacitly assumed that the transferred energy $E$
exceeds the width of spin states, given by the Korringa relaxation
rate, $\hbar/\tau_T\propto T \nu^2{\cal J}^2(T)$.

The system of equations (\ref{eq:Prates}) and (\ref{eq:kineticeq}),
with the scattering rates given by Eq.~(\ref{eq:UpsilonSR}), the
electron collision integral replaced by its average,
Eq.~(\ref{eq:Iaveraged}), and with the initial conditions for
$f(t=0,\vare,{\bm r})$, $P_J(t=0,V)$ and the boundary conditions for
$f(t,\vare,{\bm r})$, define completely the kinetics of electrons and
spins. In the stationary case ($\partial f/\partial t =0$), one may
use Eq.~(\ref{eq:Pchain}) instead of Eqs.~(\ref{eq:UpsilonSR}) and
(\ref{eq:Prates}). Note that the impurity average collision integral,
Eq.~(\ref{eq:Iaveraged}), differs from the conventional collision
integral for the electron-electron scattering. The $1/E^2$ behavior
of the kernel in Eq.~(\ref{eq:Iaveraged}) does not imply the scaling
of the distribution function found in Ref.~\onlinecite{PierreJLTP}.

The collision integral, Eq.~(\ref{eq:Iaveraged}), may be simplified
for the electron energies $\vare\gg T$. As a result, we obtain the
following kinetic equation for the distribution of ``hot'' ($\vare\gg
T$) electrons:
\begin{equation}
D\frac{\partial ^2f(\vare,x)}{\partial x^2}
= \frac{1}{\taus}
\int\chi(E)
\left\{f(\vare,x)-f(\vare+E,x)
\right\} \frac{dE}{\Tsg}.
\label{hotel}
\end{equation}
This equation is a version of a full kinetic equation, defined by
Eqs.~(\ref{eq:Prates})-(\ref{eq:Iaveraged}), which may be used for
analysis of the high energy tail of the electron distribution function
in a metal with magnetic impurities. The kernel $\chi(E)$
in Eq.~(\ref{hotel})
is asymmetric with respect to the energy transfer $E$
and for $|E|\gg\Tsg$ has the following form
\begin{equation}
\chi(E)=\frac{16}{3}\frac{\Tsg^2}{E^2}
\frac{e^{E/T}(e^{E/T}+1)}{(3e^{E/T}+1)(3+e^{E/T})}.
\label{hotelkernel}
\end{equation}
According to Eq.~(\ref{hotelkernel}) the probability
for an electron to scatter with energy gain ($E<0$) is exponentially small at
$|E|\gg T$, while the rate of scattering with an energy loss $E\gg T$
scales as a power of transferred energy,
$\propto\Tsg^2/ E^2$.

We notice that because of the relatively slow decay of
$\chi(E)$ with energy $E>0$, the
relaxation of the number of non-equilibrium electrons occurs
differently from the relaxation of their energy.
We illustrate the energy transfer from electrons to the system of
magnetic impurities by considering the following model problem.
Assume, that initially the system of electrons and magnetic impurities
is in equilibrium at temperature $T$, and then instantaneously the
electron subsystem is brought out of equilibrium, so that the new
distribution function is characterized by small deviation
$\delta f(\vare)$ from the equilibrium. The excess electron energy
$W$ per unit volume may be defined as
$W=2\nu\int \vare \delta f(\vare)d\vare.  $ The energy
$W$ will decrease in time as the result of the energy
redistribution between electron and impurity subsystems.  Eventually,
a new equilibrium with new temperature will establish.  We calculate
the reduction of the electron energy at the initial moment.  The
result is
\begin{equation}
\begin{split}
\frac{d W}{dt}=& \frac{16\Tsg}{3\taus}
\int\frac{dE}{E}\frac{e^{2E/T}-1}{(3e^{E/T}+1)(3+e^{E/T})}
\\
&
\times \int\frac{\nu \delta f(\vare)\sinh\vare/T}
{\cosh E/T+\cosh\vare/T}d\vare.
\end{split}
\label{eq:Eloss}
\end{equation}
If the distribution of electrons at the initial moment was peaked near
energy $\vare_0\gg T$, e.g.
$\nu\delta f(\vare)=\alpha\delta(\vare-\vare_0)$, then the estimate of the
energy reduction rate is
$dW/dt=\alpha\Tsg/\taus\ln\vare_0/T$. The characteristic
collision rate is $\sim 1/\taus$ and the typical energy transferred in
a collision is $\Tsg$. However, the range of the transferred energies
is broad enough to result in a logarithmic dependence on $\vare_0/T$.

In conclusion we emphasize that
Eq.~(\ref{hotelkernel}) is derived within the virial
expansion and is valid for large energy transfer. We expect that
beyond the virial expansion
the kernel $\chi(E)$ remains to be  a function of $E/\Tsg$ and $T/\Tsg$:
\begin{equation}
\chi(E)={\cal F}\left(
\frac{E}{\Tsg},\frac{T}{\Tsg}
\right).
\end{equation}
Function ${\cal F}$ characterizes the excitation spectrum of a system of
magnetic impurities at the energy scales relevant for the kinetics
of conduction electrons.
The study of its properties may provide
important information about formation of spin glass states in
metals with magnetic impurities.

\section{Discussion and Conclusions}
\label{sec:DC}

In this paper, we considered the effect of the RKKY
interaction between Kondo impurities in a metal on
kinetic properties of conduction electrons.
Specifically, we evaluated the effect of interacting magnetic
impurities on: the momentum
relaxation rate and the corresponding contribution to the Drude
resistivity of a metal $\Delta\rho_K$, the phase relaxation rate as defined by the
weak localization correction $\Delta\sigma_{\rm wl}$
to the Drude conductivity, see Eq.~(\ref{eq:deltasigmawldef}),
and the energy relaxation rate, which determines the relaxation
of non-equilibrium electrons injected into a metal.

\begin{figure}
\centerline{\epsfxsize=7cm
\epsfbox{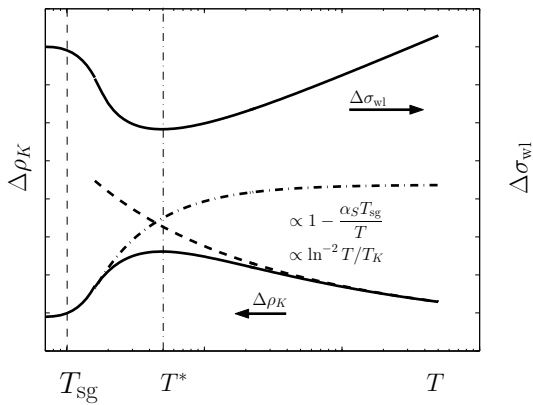}}
\caption {
  Schematic picture of the Kondo contribution to the resistivity
  $\Delta\rho_{\rm K}$ and of the weak localization correction
  $\Delta\sigma_{\rm wl}$ to the conductivity in the limit $\Tsg\gg
  \tK$.  Both $\Delta\rho_{\rm K}$ and $\Delta\sigma_{\rm wl}$ are
  expected to saturate as $T$ approaches $\Tsg$.  }
\label{fig:HTSG}
\end{figure}

The overall temperature dependence of the momentum and phase
relaxation rates differs for the cases of strong and weak RKKY
interaction between the magnetic impurities.

If the interaction $\Tsg$ between the impurities separated by a
typical distance $\ns^{-1/3}$ is strong, $\Tsg\gg \tK$, then the
momentum and phase relaxation rates are non-monotonic functions of
temperature $T$, with the maxima at $T\simeq T^*$, see
Eq.~(\ref{eq:Tstarv}) [here $\tK$ is the Kondo temperature for a
single magnetic impurity, $\ns$ is the concentration of these
impurities, and energy $\Tsg$ is defined in Eq.~(\ref{eq:Tsg})].
Therefore, if $\Tsg\gg\tK$ then one expects a maximum of the Kondo
contribution to the resistivity and a minimum of the weak localization
correction $\Delta\sigma_{\rm wl}$ at $T\simeq T^*$, see
Fig.~\ref{fig:HTSG}.  The positions of these extrema shift towards lower
temperatures with the decreasing concentration of the magnetic
impurities. At lower temperatures $T\lesssim \Tsg$
when the spin glass state is formed
both the resistivity and the WL correction to the conductivity
saturate.

\begin{figure}
\centerline{\epsfxsize=7cm
\epsfbox{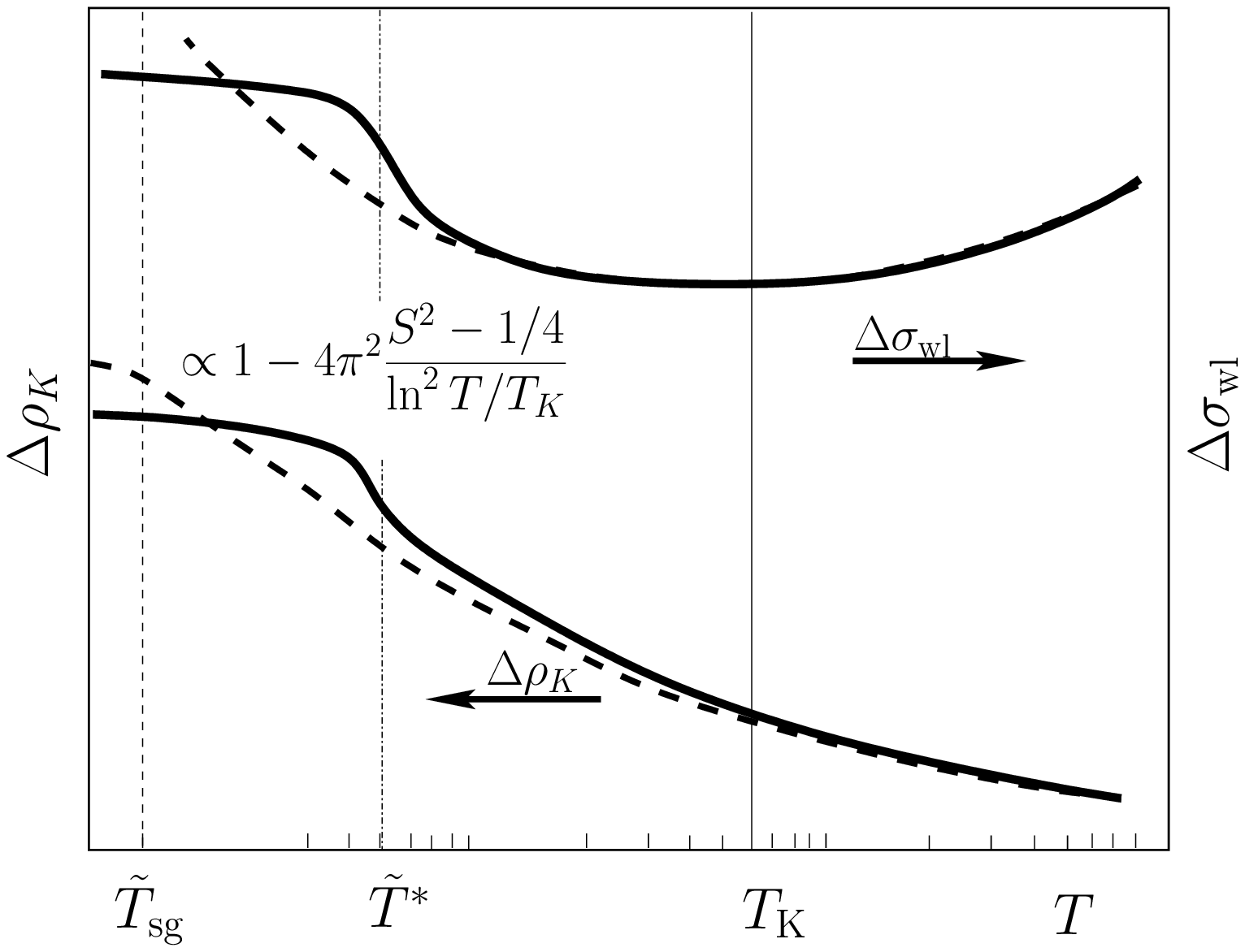}}
\caption {
  Schematic picture of the Kondo contribution to the resistivity
  $\Delta\rho_K$ and of the weak localization correction to the
  conductivity $\Delta\sigma_{\rm wl}$ in samples with low
  concentration ($\Tsg\ll\tK$) of magnetic impurities with spin
  $S>1/2$.  The solid lines represent $\Delta\rho_{\rm K}$ and
  $\Delta\sigma_{\rm wl}$ for interacting impurities. Dashed lines
  show $\Delta\rho_{\rm K}$ and $\Delta\sigma_{\rm wl}$ as if the
  interaction were absent.  }
\label{fig:HTK}
\end{figure}

In the opposite case, $\Tsg\ll \tK$, the momentum relaxation rate
increases monotonically with the decreasing $T$, and eventually
saturates at $T=0$, see the lower curve in Fig.~\ref{fig:HTK}.
The saturation level depends  on the value of impurity
spin $S$.  Thus the Kondo
contribution to the Drude conductivity is a monotonic function of
temperature. The spin-induced contribution to the phase relaxation
rate, on the contrary, has a maximum at $T\sim\tK$. If this
contribution dominates over all other mechanisms of the phase
relaxation, then the weak localization
correction to the conductivity, $\Delta\sigma_{\rm wl}$, has a minimum
at $T\sim \tK$, see the upper
curve in Fig.~\ref{fig:HTK}.
The details of the low-temperature increase of
$\Delta\sigma_{\rm wl}$ with the further reduction of temperature in
the region $T\lesssim\tK$ depend on the level of Kondo screening.  In
the case of full screening ($S=1/2$), the phase relaxation rate
vanishes at $T\to 0$, and $\Delta\sigma_{\rm wl}(T)$ diverges, see
Eq.~(\ref{eq:tauphi12}). If the screening is not complete ($S>1/2$),
then both the phase relaxation rate $1/\tilde\tau_{\rm s}$
and the weak localization correction $\Delta\sigma_{\rm wl}$
saturate at some finite level,
see Fig.~\ref{fig:HTK}. Within the
simplest model of the spin-glass state accepted in
Ref.~\onlinecite{Bergmann}, we find
\begin{eqnarray}
\Delta\sigma_{\rm wl}=\frac{e^2}{2\pi\hbar}
\sqrt{\frac{D\tilde\tau_{\rm s}}{2}},\quad
\frac{1}{\tilde\tau_{\rm s}}\simeq
\frac{8\pi\ns}{\nu}\frac{(S-1/2)^2}{\ln^2\tK/\tTsg},
\label{tausUSG}
\end{eqnarray}
where the temperature $\tTsg$ is defined in
Eq.~(\ref{eq:Tsgprime}). It is worth noting that the saturation occurs
at temperature $\sim\tilde{T}_{\rm sg}$, well below the Kondo temperature
$\tK$.

The considered limits of $\Tsg\gg\tK$ and $\Tsg\ll\tK$, and the conjecture
of Hertz~\cite{Hertz} allow us to understand the evolution of the
temperature dependence of $\Delta\rho_K(T)$ and $\Delta\sigma_{\rm
wl}$ with the concentration of impurities $\ns$. The position of
the maximum in $\Delta\rho_K(T)$ shifts continuously towards $T=0$
with the decrease of $\ns$; it reaches $T=0$ at some finite value
of $\ns$, which corresponds to $\Tsg\sim\tK$. Note that such
behavior occurs irrespective to the value of $S$. Formation of the
spin glass with $\Tsg\ll\tK$ at $S>1/2$ does not
result in the finite-temperature maximum of the function
$\Delta\rho_K$. The position of the
minimum in $\Delta\sigma_{\rm wl}$ also moves to lower values of
$T$ with the decrease of $\ns$. This shift, however, stops at
$T\sim\tK$; thus the minimum occurs at a finite temperature even
in the limit $\Tsg\ll\tK$. (Once again, here we assume that
$\taus$ is the shortest of the phase relaxation times.)

Electron scattering off interacting magnetic impurities leads to the
energy transfer from electron to the system of localized spins. The
rate of collisions with a relatively large energy transfer $E$ can be
calculated by means of the virial expansion. The corresponding full
system of kinetic equations for the electrons and spins
is derived in Section~\ref{sec:5}. The collision rate with energy loss
$E$ at
$|E|\gg{\rm max}\{\Tsg,\tK,T\}$ scales with $E$ as
$\theta (E)\Tsg/(\taus E^2)$. This asymptote of the rate
is not sensitive to the formation of the spin glass state.
However the spin glass transition affects
the electron energy relaxation for smaller energy transfers,
$|E|\lesssim \Tsg$.

In most part, the data of existing experimental
works~\cite{MohWebb,FWL,Laborde,Bauerle,larsen1,CSPWV}
can be understood within the presented here theoretical
framework.
The evolution of the
temperature dependence of the resistivity with the concentration of
magnetic impurities was studied in dilute AuFe alloys. The
investigated range of the magnetic impurity (Fe) concentration $\ns$
covered by the data of
Refs.~\onlinecite{MohWebb,FWL,Laborde,Bauerle,larsen1,CSPWV}
is extremely broad, ranging from 3.3 ppm, see
Ref.~\onlinecite{MohWebb}, up to a few percent~\cite{FWL}. For
concentrations $\ns\gtrsim 100$ ppm, the function $\Delta\rho_K(T)$
has a clear maximum~\cite{FWL,Laborde}. Its position $T_{\rm max}$
moves to lower temperatures as the impurity concentration $\ns$
decreases; the measured in Ref.~\onlinecite{Laborde} dependence of
$T_{\rm max}$ on $\ns$ is super-linear, in a qualitative agreement
with Eq.~(\ref{eq:Tstarv}).  The extrapolation of the data of
Ref.~\onlinecite{Laborde} to $T_{\rm max}=0$ yields the critical value
of Fe concentration $\approx 50$~ppm for the AuFe alloy, see also
Refs.~\onlinecite{MohWebb,CSPWV}. Finally, the observed~\cite{MohWebb}
magnetic hysteresis of the resistivity at $\ns =7$ ppm may indicate
formation of a spin glass even for such low impurity concentrations,
which is possible at $S>1/2$, see Section~\ref{sec:3B2}.

The weak localization correction to the conductivity in AuFe wires
with low impurity concentration ($\ns = 7 \sim 60$~ppm) was studied in
Refs.~\onlinecite{MohWebb,Bauerle}.  There is a proper correspondence
between the data~\cite{Bauerle} for the sample with $\ns=60$ ppm, and
the data\cite{MohWebb} for $\ns =10.9$ and $7.1$ ppm.
The values of $1/\taus$ found from the weak-localization
magnetoresistance, see Eqs.~(\ref{eq:1tausKondo}),
(\ref{eq:1tausU}) and (\ref{tausUSG}), scale
roughly linearly with $\ns$. The temperature dependence of $1/\taus$
for the investigated samples is also in accord with the theory.
Namely, the phase relaxation rate exhibits a broad plateau at
temperatures around $\tK\approx 0.3$ K (with the plateau
value\cite{Bauerle} of $1/\taus\approx 6\times 10^{10}$~s$^{-1}$
for $\ns=60$~ppm). The plateau is
followed by a decrease of this rate at lower temperatures. The
saturation\cite{Bauerle} of $1/\taus(T)$ at $T< \tK$
is compatible with the value of
impurity spins $S>1/2$.

We also notice that the data of Ref.~\onlinecite{Bauerle}
for  an AuFe alloy with $\ns=15$~ppm
are in sharp contrast with other experimental
data\cite{MohWebb,Bauerle,CSPWV,Laborde} and with the expectations
supported by the presented theory. Indeed,
Reference~\onlinecite{Bauerle}
reports the position of the Drude
resistivity maximum at $T_{\rm max}\approx 30$~mK
for the $\ns=15$~ppm sample, which is
indistinguishable from the value of $T_{\rm max}$ for the $\ns=60$~ppm
sample in the same work. This stability of $T_{\rm max}$
contradicts the dependence of $\Tsg(\ns)$, expected from other
experimental works\cite{Laborde,CSPWV} and from theory, see
Sec.~\ref{sec:3}. Also, the four-fold difference of $\ns$ between
the two samples\cite{Bauerle} resulted in a $100$--fold decrease of
the electron phase relaxation rate $1/\taus$.
This drastic change of $1/\taus$ with $\ns$ contradicts both the
quoted above measurements~\cite{MohWebb} and the theoretical
estimates, see Eqs.~(\ref{eq:1tausKondo}) and (\ref{eq:1tausU}).

Measurements of the energy relaxation in nano-wires of Au, Cu, and Ag
revealed the effect of individual magnetic
impurities~\cite{PierreJLTP}, but there was no systematic study of
the effect of RKKY interaction on the electron energy relaxation.
Measurements of the relaxation rates at energy transfers
$|E|\lesssim\Tsg$ may provide information about the excitations in a
spin glass, but we are not aware about such measurements as of yet.

We acknowledge stimulating conversations with B.L.~Altshuler.
We are grateful to C.~Bauerle, N.~Birge, J.~Mydosh, H.~Pothier,
L.~Saminadayar for useful comments and illuminating discussions
of experiments, and to C. Chamon, L.~Cugliandolo, L.~Ioffe for discussions
of spin glass phenomena. This work was supported by
NSF grants DMR97-31756, DMR01-20702, DMR02-37296, and EIA02-10736.

\appendix

\section{Matrix elements $A_{JJ'}$}
\label{app:A}

In this Appendix we present the expression for the $A_{JJ'}$ factors in
Eq.~(\ref{eq:deltaK}). First we notice that the convenient form to calculate
$\langle JM|\hat S_\alpha|J'M'\rangle $ is to use the basis of spin states for
two independent spins: $|m_1m_2\rangle$, where $m_1$ and $m_2$ are the spin
components along some direction. We have
\begin{eqnarray}
\eta_\alpha(JMJ'M')& = &
\langle JM|\hat S_\alpha|J'M'\rangle
\nonumber
\\
& = &
\!\!\!\!
\sum\limits_{m_1,m_2,m_1'}\!\!\!\!\!
C_{m_1m_2}^{JM}\langle m_1|\hat S_\alpha|m_1'\rangle
C_{m_1'm_2}^{J'M'},
\nonumber
\end{eqnarray}
where the Clebsh-Gordon coefficients are expressed in terms of the Wigner
$3j-$symbols as
\begin{eqnarray}
C_{m_1m_2}^{JM}=\sqrt{2J+1}
\left(
\begin{array}{ccc}
  J  & S   & S \\
  -M & m_1 & m_2 \\
\end{array}%
\right).
\end{eqnarray}

The matrix element $A_{JJ'}$ may be represented in terms of
$\eta_\alpha$:
\begin{equation}
A_{JJ'}=\sum_{\alpha=x,y,z}\sum_{M'}\eta^2_\alpha(JMJ'M').
\end{equation}
We find that
\begin{subequations}
\label{eq:AJJ-explicit}
\begin{eqnarray}
A_{J,J+1} & = &
\frac{(J+1)(2S-J)(2S+J+2)}{4(2J+1)},
\\
A_{J,J} & = & \frac{J(J+1)(2J+1)}{4(2J+1)},
\\
A_{J,J-1} & = &
\frac{J(2S-J+1)(2S+J+1)}{4(2J+1)},
\end{eqnarray}
\end{subequations}
and all other elements vanish. From Eqs.~(\ref{eq:AJJ-explicit})
we verify explicitly that
\begin{equation}
\sum_{J,J'=0}^{2S}(2J+1)A_{JJ'}=S(S+1)(2S+1)^2.
\label{eq:sumrule}
\end{equation}

\begin{figure}
\centerline{\epsfxsize=7cm
\epsfbox{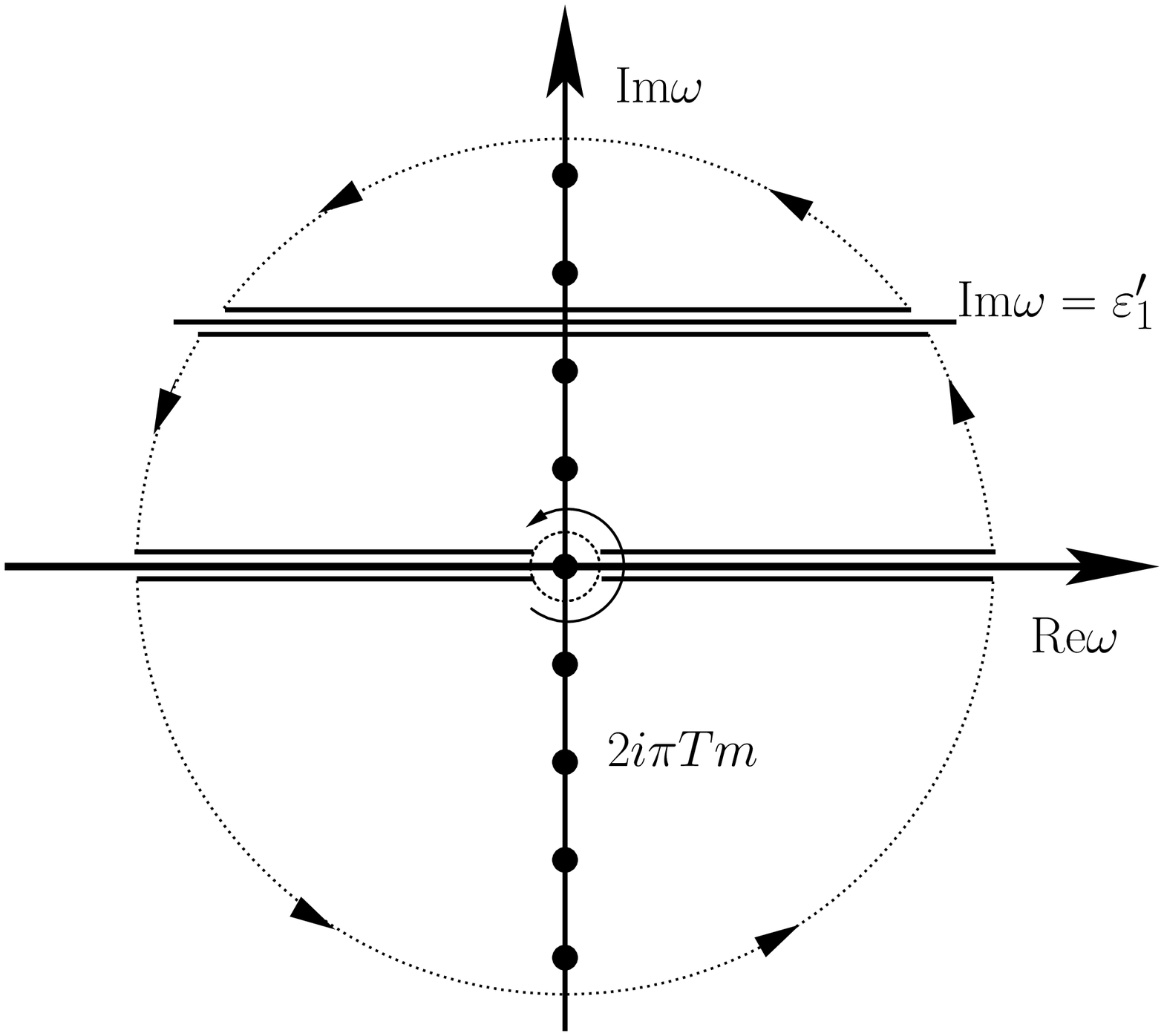}}
\caption {The contour for calculation of the $T-$matrix,
Eq.~(\ref{eq:TMatsub1}), in the Matsubara representation.
The contributions from the dotted parts of the contour
vanish. }
\label{fig:contour1}
\end{figure}

\section{Electron $T-$matrix.}
\label{app:B}

For completeness of the presentation, we show how the imaginary part of
the electron $T-$matrix may be related to the spin correlation
function, defined by Eq.~(\ref{eq:Kdef}).
In the Matsubara representation the $T-$matrix
is given by the following expression:
\begin{equation}
{\cal T}(i\vare_n)={\cal J}^2 T\sum_{\omega_m}
\int\frac{d\bfp}{(2\pi)^3}
{\cal K}(i\omega_m){\cal G}(i(\vare_n-\omega_m),\bfp),
\label{eq:TMatsub}
\end{equation}
where ${\cal K}(i\omega_m)$ is the Fourier component at $\omega_m=2\pi m T$
of the
Matsubara spin correlator
\begin{equation}
{\cal K}(\tau)=\sum_{\xi\xi'}\rho_\xi e^{-(E_\xi-E_{\xi'})\tau}
\left|
\langle \xi | {\bm S} | \xi'\rangle
\right|^2
\label{eq:Kmatsubara}
\end{equation}
[compare to Eqs.~(\ref{eq:Kdef}) and (\ref{eq:Kgeneral})
for the real time spin correlator $K(t)$], and
\begin{equation}
{\cal G}(i\vare_n,\bfp)=\frac{1}{i\vare_n-\xi(\bfp)-\Sigma(\vare_n)}
\end{equation}
is the electron Green's function at Matsubara frequency
$\vare_n=\pi(2n+1)T$ [compare to Eq.~(\ref{eq:Gra})].
The integration over momentum in
Eq.~(\ref{eq:TMatsub}) gives
\begin{equation}
{\cal T}(i\vare_n)= -i \pi\nu {\cal J}^2
\sum_{\omega_k}{\cal K}(i\omega_m){\rm sign}(\vare_n-\omega_m).
\label{eq:TMatsub1}
\end{equation}
Next we perform the standard procedure of the analytical continuation
in Eq.~(\ref{eq:Kmatsubara}).
We replace the sum
over discreet $\omega_m $ by the integral over complex
$\omega$:
\begin{equation}
T\sum_{\omega_m} {\cal F}(i\omega_m) =
i \int_C \frac{d\omega}{4\pi}\coth \frac{\omega}{2T}{\cal F}(\omega).
\label{eq:TMatsubcont}
\end{equation}
This procedure is valid for an arbitrary function ${\cal F}$,
analytic inside the contour $C$ of integration. We choose $C$ to be
a circle of infinite radius with two cuts at
${\rm Im}\omega=0$ and ${\rm Im}\omega=\vare_n$. Inside the contour of the
integration, the function ${\cal K}(\omega)$ is
analytic and only the poles of $\coth\omega/2T$ contribute to the integral.
Neglecting the contribution from the pieces of the circle, and keeping the
contribution along the cuts, we obtain
\begin{equation}
\begin{split}
& \frac{{\cal T}(i\vare_n)} {\pi\nu {\cal J}^2} =
- \int\limits_{-\infty}^{+\infty}\!\!
\tanh \frac{\omega}{2T} {\cal
K}(i\vare_n+\omega)\frac{d\omega}{2\pi}
\\
& +
 {\rm sign}\vare_n
\!\!\!
\int\limits_{-\infty}^{+\infty}\!\!\!
\coth \frac{\omega}{2T}
\frac{
{\cal K}(\omega+i0)-
{\cal K}(\omega-i0)}{2}\frac{d\omega}{2\pi}.
\end{split}
\label{eq:B6}
\end{equation}
The imaginary part of the $T-$matrix is given by
\begin{equation}
{\rm Im}T(\vare)= i \frac{{\cal T}(\vare+i0)-{\cal
T}(\vare-i0)}{2}.
\label{eq:B7}
\end{equation}
We substitute Eq.~(\ref{eq:B6}) in Eq.~(\ref{eq:B7}),
take into account  relation
\begin{equation}
K(\omega)=
 i \frac{{\cal K}(\omega+i0)-{\cal K}(\omega-i0)}{e^{\omega/T}-1}
\label{eq:KrealMat}
\end{equation}
between the real time spin
correlator, Eq.~(\ref{eq:Kdef}), and the Matsubara spin
correlator, Eq.~(\ref{eq:Kmatsubara}), and obtain Eq.~(\ref{eq:ImT}).

\section{Cooperon Self Energy}
\label{app:C}

\begin{figure}
\centerline{\epsfxsize=7cm
\epsfbox{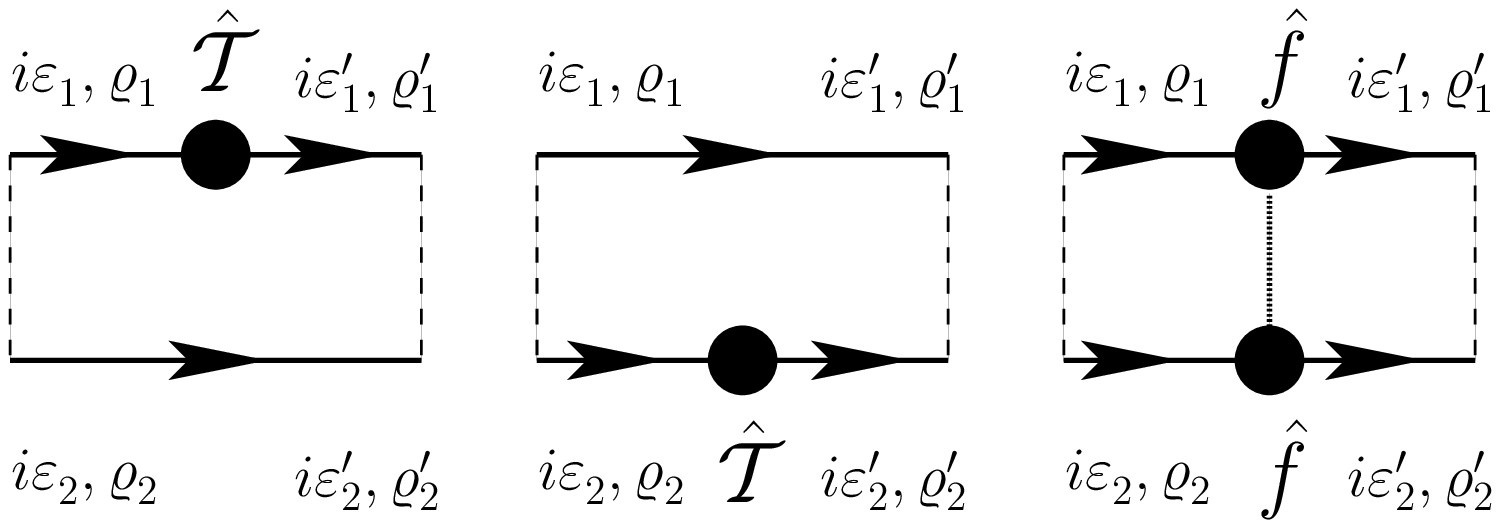}}
\caption {The left two diagrams represent the self-energy
corrections to the Cooperon and contain the single electron
$T-$matrix. The diagram on the right is the vertex correction and
is related to the two electron counterpart of a $T-$matrix.
 }
\label{fig:CoopSE}
\end{figure}

The Cooperon self energy $\Sigma_{\rm rr}(\cdot)$
may be obtained as a result of the analytical continuation of
$\Sigma\left(
\begin{matrix}
i\vare_1, & i\vare'_1 \\
i\vare_2, & i\vare'_2
\end{matrix}
\right)$,
written in the Matsubara representation:
\begin{equation}
\Sigma^{\rm rr}_s\left(
\begin{matrix}
\vare_1, & \vare'_1 \\
\vare_2, & \vare'_2
\end{matrix}
\right)=
\Sigma_s\left(
\begin{matrix}
\vare_1+i0, & \vare'_1+i0 \\
\vare_2-i0, & \vare'_2-i0
\end{matrix}
\right).
\label{eq:Sigmarr}
\end{equation}
The Cooperon self energy is related to the scattering
matrix $\cal S$ off a magnetic impurity for two electrons in a
singlet state:
\begin{equation}
\Sigma_s(\cdot)=2\pi \nu \ns \overline{{\cal S}(\cdot)}.
\label{eq:C2}
\end{equation}
In Eq.~(\ref{eq:C2})  ${\cal S}$ is averaged over
various magnetic impurities. The singlet component
${\cal S}_s$ is related to the full matrix $\hat {\cal S}$ of two electron
scattering:
\begin{equation}
{\cal S}_{s}
\left(
\cdot
\right)=
\frac{1}{2}\left(
{\cal S}^{\uparrow\uparrow}_{\downarrow\downarrow}(\cdot)+
{\cal S}_{\uparrow\uparrow}^{\downarrow\downarrow}(\cdot)-
{\cal S}_{\uparrow\downarrow}^{\downarrow\uparrow}(\cdot)-
{\cal S}^{\uparrow\downarrow}_{\downarrow\uparrow}(\cdot)
\right).
\label{eq:SigmaSinglet}
\end{equation}
Matrix ${\cal S}$, in its turn, contains the contribution from
three diagrams, shown in Fig.~\ref{fig:CoopSE}, and has the
following form:
\begin{equation}
\begin{split}
{\cal S}&^{\varsigma_1^{}\varsigma_1'}_{\varsigma_2^{}\varsigma_2'}
\left(
\begin{matrix}
i\vare_1, & i\vare'_1 \\
i\vare_2, & i\vare'_2
\end{matrix}
\right) =
{\cal V}^{\varsigma_1^{}\varsigma_1'}_{\varsigma_2^{}\varsigma_2'}
\left(
\begin{matrix}
i\vare_1, & i\vare'_1 \\
i\vare_2, & i\vare'_2
\end{matrix}
\right)
\\
&-
\frac{{\cal T}(i\vare_1)-{\cal T}(i\vare_2)}{2\pi i\nu}
\theta(-\vare_1\vare_2) {\rm sign}\vare_1
\delta_{\vare_1\vare_1'}\delta_{\vare_2\vare_2'}
\delta_{\varsigma_1^{}\varsigma_1'}\delta_{\varsigma_2^{}\varsigma_2'}.
\label{eq:CoopSEMat}
\end{split}
\end{equation}
Here the second term contains the single electron $T-$matrix
considered in Appendix \ref{app:B}. Below we focus on the
irreducible component of the scattering matrix of two electrons in
a singlet state, represented by
\begin{equation}
\begin{split}
{\cal V}& ^{\varsigma_1^{}\varsigma_1'}_{\varsigma_2^{}\varsigma_2'}
\left(
\begin{matrix}
i\vare_1, & i\vare'_1 \\
i\vare_2, & i\vare'_2
\end{matrix}
\right)
= \theta(-\vare_1\vare_2)\theta(-\vare'_1\vare'_2)
\sum_{\xi\xi'}\frac{e^{-E_\xi/T}}{Z}
\\
&\times
T\sum_{\omega_m}f^{\xi\xi'}_{\varsigma_1^{}\varsigma_1'}(i\omega_m)
f^{\xi\xi'}_{\varsigma_2^{}\varsigma_2'}(-i\omega_m)
\delta_{\vare_1+\omega_m,\vare_1'}\delta_{\vare_2,\vare_2'+\omega_m},
\label{eq:CoopSEMat1}
\end{split}
\end{equation}
where $f^{kk'}_{\varsigma^{}\varsigma'}(i\omega_m)$ is the scattering
amplitude of electrons with initial spin state $\varsigma'$ to the spin state
$\varsigma$, accompanied with the change of the state of magnetic impurities
from $\xi'$ to $\xi$. The summation over $\xi$ and $\xi'$
runs over all possible states of the
impurity spin and $Z=\sum_\xi e^{-E_\xi/T}$.
For scattering amplitudes
in the Born approximation we have
\begin{equation}
f^{\xi\xi'}_{\varsigma\varsigma'}=
{\cal J} \langle k|\hat {\bm S}| k'\rangle
\hat{\bm \sigma}_{\varsigma\varsigma'}.
\label{eq:fexp}
\end{equation}

We substitute Eq.~(\ref{eq:fexp}) into Eq.~(\ref{eq:CoopSEMat1}),
and use Eq.~(\ref{eq:SigmaSinglet}) to write down the singlet
component of the vertex part of the Cooperon ${\cal S}$
matrix:
\begin{eqnarray}
{\cal V}_s\left(
\begin{matrix}
i\vare_1, & i\vare'_1 \\
i\vare_2, & i\vare'_2
\end{matrix}
\right)  & = &  {\cal J}^2 \theta(-\vare_1\vare_2)\theta(-\vare'_1\vare'_2)
T\sum\limits_{\omega_m} \delta {\cal K}_2(i\omega_m)
\nonumber
 \\
 && \times
\delta_{\vare_1+\omega_m,\vare_1'}\delta_{\vare_2,\vare_2'+\omega_m}.
\label{eq:CoopSEsinglet}
\end{eqnarray}

\begin{figure}
\centerline{\epsfxsize=7cm
\epsfbox{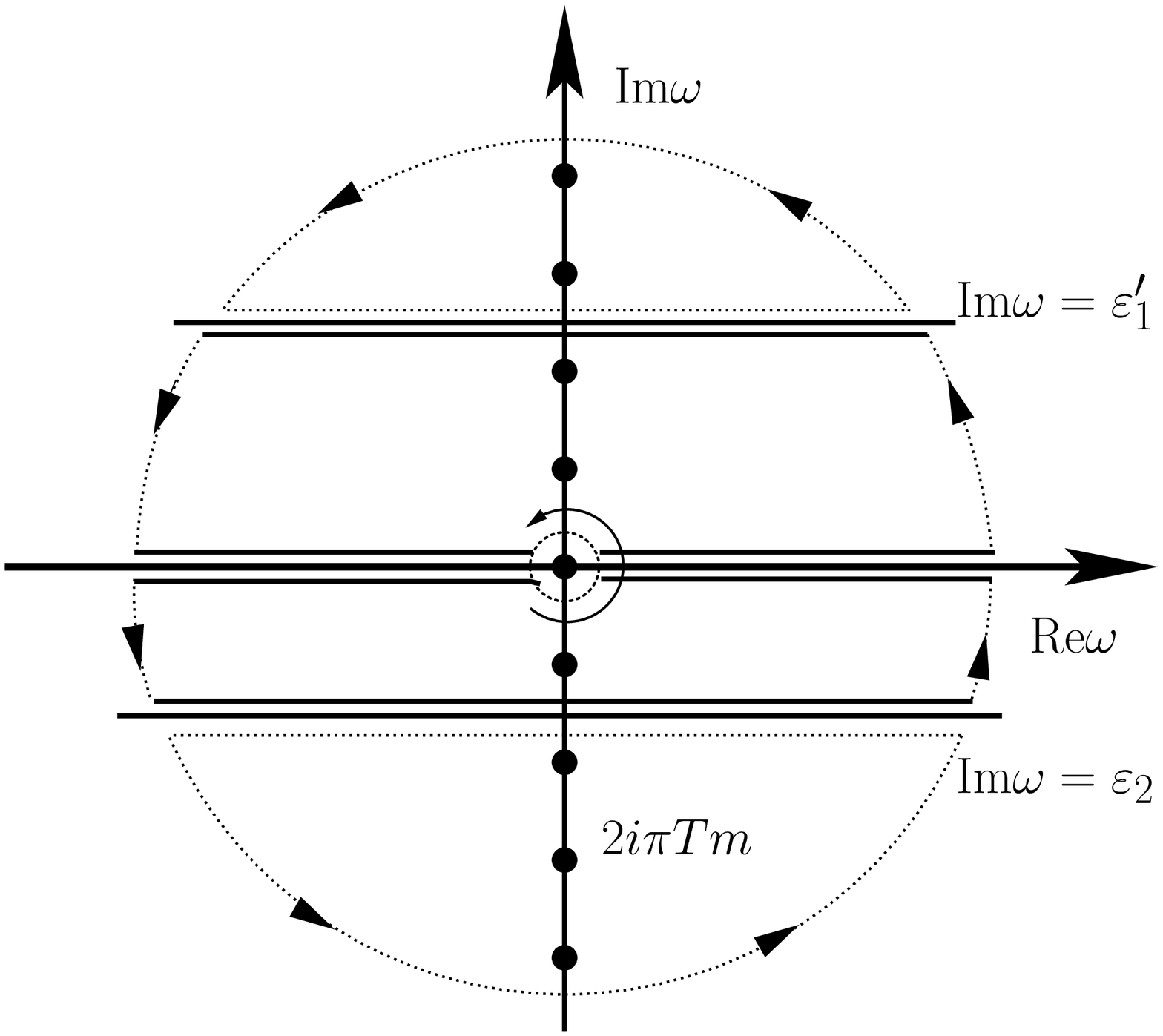}}
\caption {
The contour for calculation of the vertex correction
$\hat {\cal V}_s$ to the Cooperon. The contributions from
the dotted parts of the contour vanish. }
\label{fig:contourCoop}
\end{figure}

Next we perform the analytical continuation of ${\cal V}_s$.
We consider the case $\vare_1'>0$ and $\vare_2<0$, see Eq.~(\ref{eq:Sigmarr}).
Using Eq.~(\ref{eq:TMatsubcont}), we replace the sum
over the Matsubara frequencies, $\omega_m$, by the integral over a contour in
the complex plane. In the present case the contour of the integration is shown
in Fig. \ref{fig:contourCoop},
and contains three cuts: ${\rm Im}\omega = \vare_2,\ 0,\ \vare_1'$.
We notice, that the contour parts above the upper cut and
below the lower cut do not contribute to the integral. The remaining parts of the
contour
along the cuts after the continuation $i\vare_1'\to \vare_1'+i0$ and
$i\vare_2\to \vare_2-i0$ give
\begin{equation}
\begin{split}
{\cal V}_s &
\left(
\begin{matrix}
\vare_1+i0, &  \vare'_1+i0 \\
\vare_2-i0, &  \vare'_2-i0
\end{matrix}
\right)\! =\! \!\!\! \int\limits_{-\infty}^{+\infty}\!\!
d\omega
\delta(\vare_1+\omega-\vare_1')\delta(\vare_2-\omega-\vare_2')
\\
&\times 2\pi {\cal J}^2 \Big\{
\coth\frac{\omega}{2T}[\delta {\cal K}_2(\omega+i0)-
\delta {\cal K}_2(\omega-i0)]
\\
&-
\tanh\frac{\omega}{2T}
[\delta {\cal K}_2(\omega+\vare_1'+i0)-
\delta {\cal K}_2(\omega+\vare_2-i0)]
\Big\}.
\label{eq:Vs}
\end{split}
\end{equation}
This expression may be further simplified in the case when
$\vare_1'=\vare_2$ with the help of Eq.~(\ref{eq:KrealMat}).

\section{Calculations of the virial correction to $\Delta\sigma_{\rm wl}$}
\label{app:D}

To calculate the WL correction to the conductivity, we substitute
Eq.~(\ref{eq:CoopVir}) with the self energy defined by
Eq.~(\ref{eq:SEvirial}) into Eq.~(\ref{eq:wlgeneral}). As a result
we obtain
\begin{equation}
\Delta\sigma_{\rm wl} =
\Delta\sigma^{(0)}_{\rm wl}+\Delta\sigma^{(1,a)}_{\rm wl}
+\Delta\sigma^{(1,b)}_{\rm wl}.
\label{eq:D1}
\end{equation}
Here $\Delta\sigma^{(0)}_{\rm wl}$ is given by
Eq.~(\ref{eq:wl1d0}), and the second term is
\begin{subequations}
\begin{eqnarray}
\Delta\sigma^{(1,a)}_{\rm wl}
& = & -\frac{\Delta\sigma^{(0)}_{\rm wl}}{4}
\int
\frac{\delta\overline{K_2(\omega)}}{S(S+1)}
\frac{\omega}{T}\frac{1}{1-e^{-\omega/T}}
\frac{d\omega}{\pi}
\nonumber
\\
& = &\frac{\Delta\sigma^{(0)}_{\rm wl}}{2} \frac{\alpha_S\Tsg}{T},
\end{eqnarray}
with $\alpha_S$ defined by Eq.~(\ref{eq:alphaS}), see also
Table~\ref{tab:1}. The third term in Eq.~(\ref{eq:D1}) is
\begin{equation}
\Delta\sigma^{(1,b)}_{\rm wl}
= -\frac{\Delta\sigma^{(0)}_{\rm wl}}{2}
\int
\frac{\delta\overline{K_2(\omega)}}{S(S+1)}
\frac{\omega}{T}\frac{\varphi(\omega)}{1-e^{-\omega/T}}
\frac{d\omega}{\pi},
\end{equation}
\end{subequations}
where
$$
\varphi(\omega)= \frac{1}{4}-
\frac{{\rm Im} \sqrt{1+i\omega\taus/2}}
{\omega\taus\sqrt{1+\omega^2\taus^2/4}}.
$$
As $\omega\taus\to 0$ function $\varphi(\omega)$ vanishes. Therefore,
at low temperature $T\taus\ll 1$ only the second term
$\Delta\sigma^{(1,a)}_{\rm wl}$
in Eq.~(\ref{eq:D1}) remains. In the opposite limit,
$T\taus\gg 1$, we use the following property of $\varphi(\omega)$
$$
\int_{-\infty}^{+\infty} \frac{\varphi(\omega)}{\omega^2}d\omega
=\frac{3\pi}{32}\taus
$$
and obtain
\begin{equation}
\Delta\sigma^{(1,b)}_{\rm wl}\approx
\frac{\pi}{120}\frac{(4S+1)(4S+3)}{2S+1}
\Delta\sigma^{(0)}_{\rm wl} \Tsg\taus,
\quad T\taus\gg 1.
\end{equation}


\begin{thebibliography}{99}

\bibitem{Pierre} F.~Pierre, A.B.~Gougam, A.~Anthore, H. Pothier, D. Esteve, and
N.O.~Birge, preprint cond-mat/0302235.

\bibitem{PierreJLTP} F.~Pierre, H. Pothier, D. Esteve, and
M.H.~Devoret, Journal of Low Temp. Phys. {\bf 118}, 437 (2000).

\bibitem{KG} A. Kaminski and L.I. Glazman, Phys. Rev. Lett. {\bf 86},
  2400 (2001).

\bibitem{Kondomin} W.J. de Haas, J.H.~de Boer, and
G.J.~van der Berg, Physica {\bf 1}, 1115 (1934).

\bibitem{Laborde} O.~Laborde and P. Radhakrishna,
Solid State Commun. {\bf 9}, 701 (1971).

\bibitem{Ford} P.J.~Ford and J.A.~Mydosh, Phys. Rev. B {\bf 14},
2057 (1976).

\bibitem{mydosh} J.A.~Mydosh, {\it Spin Glasses}
(Taylor \& Francis, London, 1993).

\bibitem{larsen} U.~Larsen, Phys. Rev. B {\bf 14}, 4356 (1976).

\bibitem{larsen1} J.S.~Schilling, P.J.~Ford, U.~Larsen, and
J.A.~Mydosh, Phys. Rev. B {\bf 14}, 4368 (1976).

\bibitem{MohWebb} P. Mohanty and R.A.~Webb, Phys. Rev. Lett. {\bf 84}, 4481
(2000).

\bibitem{Bauerle} F.~Schopfer, C.~B\"auerle, W.~Rabaud, and
L.~Saminadayar, Phys. Rev. Lett. {\bf 90}, 056801 (2003).

\bibitem{BFK} A.A. Bobkov, V.I. Fal'ko, and D.E. Khmel'nitskii,
Zh. Exp. Teor. Fiz. {\bf 98}, 703 (1990) [Sov. Phys. - JETP {\bf 71},
393 (1990)]; V.I. Fal'ko,  J. Phys.: Condens. Matter {\bf 4}, 3943 (1992).

\bibitem{VG} M.G. Vavilov and L.I. Glazman, Phys. Rev. B {\bf 67},
115310
(2003).

\bibitem{Goppert} G. G\"oppert, Y.M. Galperin, B.L. Altshuler, and H. Grabert,
Phys. Rev. B {\bf 66}, 195328 (2002).

\bibitem{KondovsMagField} see, e.g. A.C.~Hewson, {\it The Kondo
problem to Heavy Fermions} (Cambridge University Press, Cambridge, UK,
1997).

\bibitem{LKh} A.I. Larkin, and D.E. Khmelnitsky,
Zh. Exp. Teor. Fiz. {\bf 58}, 1789 (1970)
[Sov. Phys. JETP {\bf 31}, 958 (1970)].

\bibitem{LMK} A.I. Larkin, V.I. Melnikov and D.E. Khmelnitsky,
Zh. Exp. Teor. Fiz. {\bf 60}, 846 (1971)
[Sov. Phys. JETP {\bf 33}, 458 (1971)].

\bibitem{HLN} S. Hikami, A.I. Larkin and Y. Nagaoka, Progr. Theor. Phys.
{\bf 63}, 707 (1980).

\bibitem{WL} B.L.~Altshuler, A.G.~Aronov, D.E.~Khmelnitsky,
A.I.~Larkin in
{\it Quantum Theory of Solids} (Mir publisher, Moscow, 1982);
B.L. Altshuler, A.G. Aronov, M.E. Gershenson, and Yu.V. Sharvin,
  Sov. Sci. Rev. A.: Phys. {\bf 9},
223 (Harwood Academic Publishers, 1987).

\bibitem{BA} V. Barzykin and I. Affleck, Phys. Rev. B {\bf 61}, 6170
(2000).

\bibitem{GL} V.M.~Galitski and A.I.~Larkin, Phys. Rev. B {\bf 66},
064526 (2002).

\bibitem{FischerHertz} K.H.~Fischer and J.~Hertz, {\it Spin Glasses}
(Cambridge University Press, New York, 1993).

\bibitem{AL} I. Affleck and A.W.W.~Ludwig, Phys. Rev. B {\bf 48},
7297 (1993).

\bibitem{NB} P. Nozi\'eres and A. Blandin, J. Physique {\bf 41}, 193 (1980).

\bibitem{Varma} B. A. Jones and C. M. Varma, Phys. Rev. Lett. {\bf 58},
843 (1987).

\bibitem{Hertz} J.A.~Hertz, Phys. Rev. B {\bf 19}, 4796 (1979).

\bibitem{AAK} B.L.~Altshuler, A.G.~Aronov, and D.E.~Khmelnitsky,
J. Phys. C: Solid State Phys. {\bf 15}, 7367 (1982).

\bibitem{AAG} I.L.~Aleiner, B.L.~Altshuler, and M.E. Gershenson,
Waves in Random Media {\bf 9},  201  (1999).


\bibitem{Bergmann} W. Wei, G. Bergmann, and R.–P. Peters,
Phys. Rev. B {\bf 38}, 11751 (1988).



\bibitem{FWL} P.J.~Ford, T.E.~Whall, and J.W.~Loram, Phys. Rev. B
{\bf 2}, 1547 (1970).

\bibitem{CSPWV} V.~Chandrasekhar, P.~Santhanam, N.A.~Penebre,
R.A.~Webb, H.~Vloeberghs, C.~Van~Haesendonck, and Y.~Bruynseraede,
Phys. Rev. Lett. {\bf 72}, 2053 (1994).

\end{thebibliography}
\end{document}